\documentclass[final]{IEEEtran}
\usepackage{amsthm,amssymb,graphicx,multirow,amsmath,color,amsfonts}%
\usepackage[update,prepend]{epstopdf}
\usepackage[noadjust]{cite}
\usepackage[latin1]{inputenc}
\usepackage{tikz}
\usepackage{bbm} %
\usepackage{pdfpages}
\usepackage{upgreek}
\usepackage{multirow}
\usepackage{comment}
\usepackage{tabularx}
\usepackage{inputenc}
\usepackage{colortbl}
\usepackage{xcolor}
\usepackage{gensymb}

\usepackage{lettrine}
\usepackage{subfig}

\begin{document}
\graphicspath{{./Figures/},{plots/pdf/}}
\title{
A Schottky-Diode-Based Wake-Up Receiver for IoT Applications
}
\author{
Mahmoud Elhebeary, Student Member, IEEE, Samer Hanna, Student Member, IEEE, Sudhakar Pamarti, Senior Member, IEEE, Danijela Cabric, Senior Member, IEEE, and Chih-Kong Ken Yang, Fellow, IEEE
\thanks{Mahmoud Elhebeary, Samer Hanna, Sudhakar Pamarti, Danijela Cabric, and Chih-Kong Ken Yang are with the  University of  California, Los Angeles, CA 90095 USA. Email: \{mahmoudrashad, samerhanna, spamarti, danijela, yangck\}@ucla.edu} %
}
\maketitle
\begin{abstract}
This paper presents an always-on low-power wake-up receiver (WuRx) that activates the remainder of the system when a wake-up signal is detected. The proposed receiver has two phases of waking up. The first phase uses an integrated CMOS Schottky diodes to detect the signal power at a low bias current. The approach dissipates low quiescent power and allows the reuse of the design in multiple frequency bands with only modifying the matching network. In the second phase, a data-locked startable oscillator is proposed to correlate the received data with a target signature. This design eliminates the area and power dissipation of an external crystal oscillator and only turns on when the second phase is activated.  By correlating to a target signature, the second phase also reduces the probability of a false alarm (PFA) that would otherwise wake up the high-power bulk of the system. The two-phase approach leads to significant reduction in average power consumption when compared to a single-phase design.
This implementation targets sub-ms wake-up latency and operates in the unlicensed band at a 750-MHz carrier frequency with a data rate of 200 kbps. The design achieves $\sim$8.45pJ/bit and $<$-50 dBm of input sensitivity and average power of 1.69$\mu$W. The system is implemented in 65-nm CMOS technology and occupies an area of 1mm$\times$0.75mm.
\end{abstract}
\section{Introduction} \label{sec:intro}
\lettrine[lines=2]{I}{nternet}-of-Things (IoT) has enabled a wide array of applications by interconnecting a multitude of leaf nodes. In such systems, leaf nodes collect data from their environment and transmit the data to a gateway, which in turn stores and processes the data in a cloud or database. Leaf nodes, especially ones that are widely disseminated without connections to the power grid, have a limited power budget which constrains how often each leaf node can communicate with its gateway. In environmental, biomedical, or agricultural applications, a mobile gateway \cite{1-1-3}-\cite{1-3-4} that collects the data may only be periodically available. Always-on wake-up receivers (WuRx) have been introduced to constantly monitor a frequency band of interest at a very low power level and then awaken a primary transceiver of the IoT system upon receiving a wake-up signature. In applications such as a wireless body area network (WBAN), a WuRx should have low power dissipation of $<$5$\mu$W due to the high cost of battery replacement, moderate data rates of 100's of kbps to handle modest system configuration data, and sensitivity of $<$-40Bm. We further designed our target system to a wide data bandwidth to enable low wake-up latency of $<$200$\mu$s so that the leaf node can quickly link with an available gateway. This paper introduces several circuit techniques to lower the power and reduce the energy-per-bit to $<$10 pJ/bit. Because a WuRx operates in an unlicensed frequency band, multiple interferes can be present. False alarms can result in a large power penalty by waking up the system unnecessary. This paper further explores the power impact of a two-phase wake-up that incorporates digital correlation to reduce false alarms.  
\par
\begin{figure}[t!]
	\centering
	\includegraphics[scale=0.2]{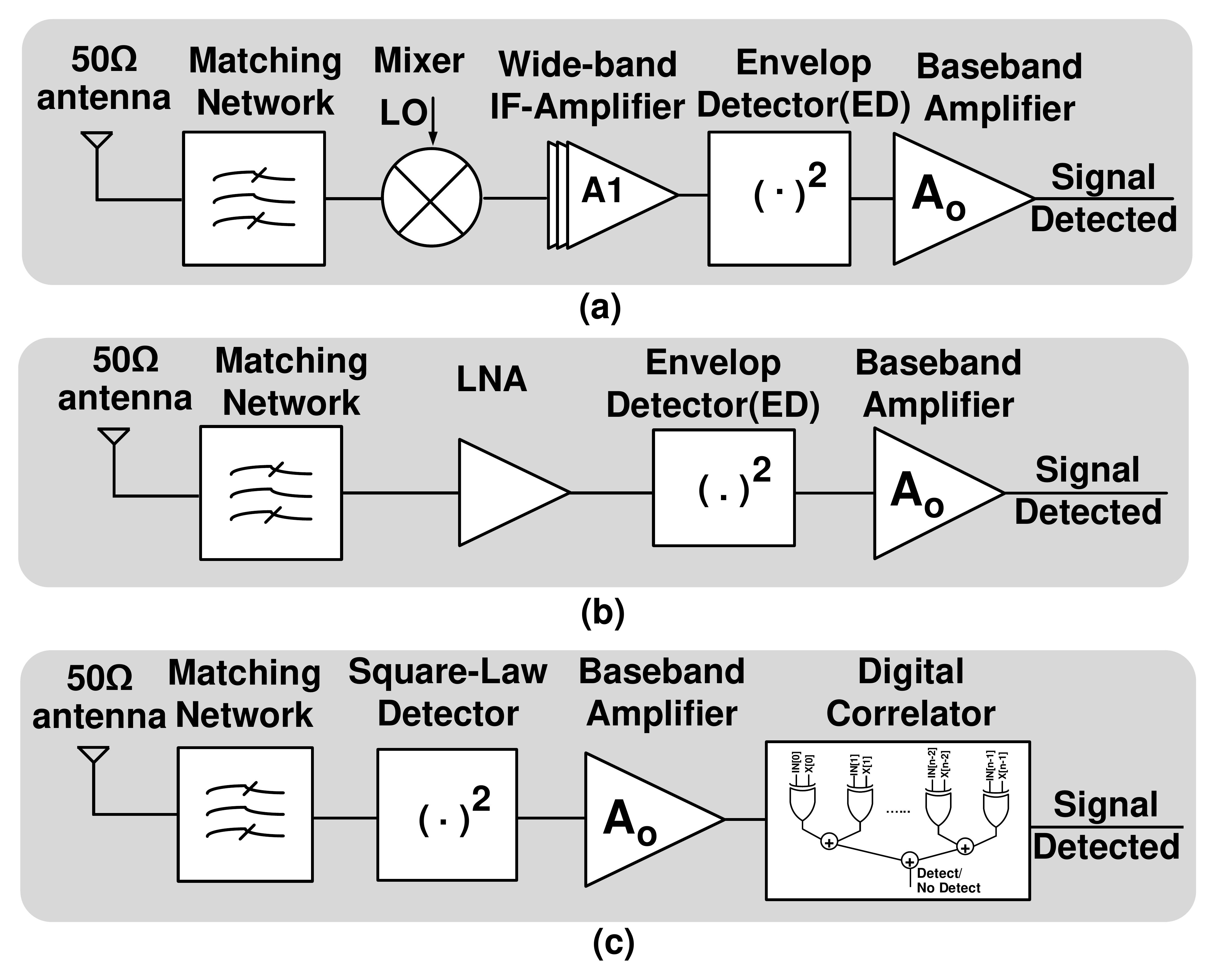}
	\caption{Architecture of (a) a conventional IF/Uncertain-IF Superhetrodyen WuRx, (b) a conventional tuned RF WuRx, and (c) a square-law detector based WuRx.}
	\label{fig:2-1}
\end{figure}

Always-on WuRx can be categorized into three main architectures dictated by their front end as shown in Fig. \ref{fig:2-1}. The first architecture is based on superheterodyne mixers  \cite{1-1-3}-\cite{1-1-8} which achieves good interference rejection as well as the required fast start-up time requirement. However, the high power consumed in the local oscillator and mixer power results in high energy per bit. The second architecture utilizes a frontend with tuned-RF filter and LNA \cite{1-2-1}-\cite{1-2-3} which achieves high sensitivity at the expense of the power consumption of the LNA. The third architecture eliminates the mixer and LNA and uses a square-law detector as an energy detector \cite{1-3-1}-\cite{1-3-7}. The envelop detector has the same structure as the square-law and are interchangeable, however we the term ED for large signal detection whereas for small signal detection we use square-law or energy detector. This architecture can be designed with high-latency \cite{1-3-4}-\cite{1-3-7} and low data rate ($<$1kbps) which in turn results in low power and high sensitivity due to the reduced bandwidth. Alternatively, designs targeting low-latency applications \cite{2-1-2} have also been shown to run at high data rates ($>$50kbps) and moderate sensitivity.A digital correlator would benefit the 3 architectures, however it would have maximum benefit when applied for square-law detector based architecture 
\par
In this work, we propose a two-phase always-on WuRx that can be used in different bands. In the first phase, as the energy detector, we propose using Schottky diodes. Through a fully-integrated solution, we can minimize parasitic capacitance allowing operation at a high-frequency range \cite{1-4-1}. The second phase utilizes a novel data-locked startable oscillator which allows digital data correlation while avoiding the need for an external clock or external crystal oscillator. A digital correlator compares the received data to a preset signature to avoid falsely activating the primary transceiver by interferers. By optimizing the correlator length and the energy detector thresholds, we show that the two-phase WuRx can reduce the system-level energy consumption by up to 60x compared to the equivalent single-stage energy detector. A wide-bandwidth front-end is adopted in the design to satisfy the wake-up latency of 200$\mu$s corresponding to a signature of 40-bits at 200kbps data rate. 
\par
The paper is organized as follows.  Section \ref{sec:model} shows the proposed system architecture, the system modeling, and the trade-offs of design parameters. Section \ref{sec:sysimp} discusses the circuit implementation to achieve targeted sensitivity and lower overall power consumption. The measurement results for the proposed WuRx is discussed in Section \ref{sec:MeasRes}.

\begin{figure*}[!t]
	\centering
	\includegraphics[scale=0.7]{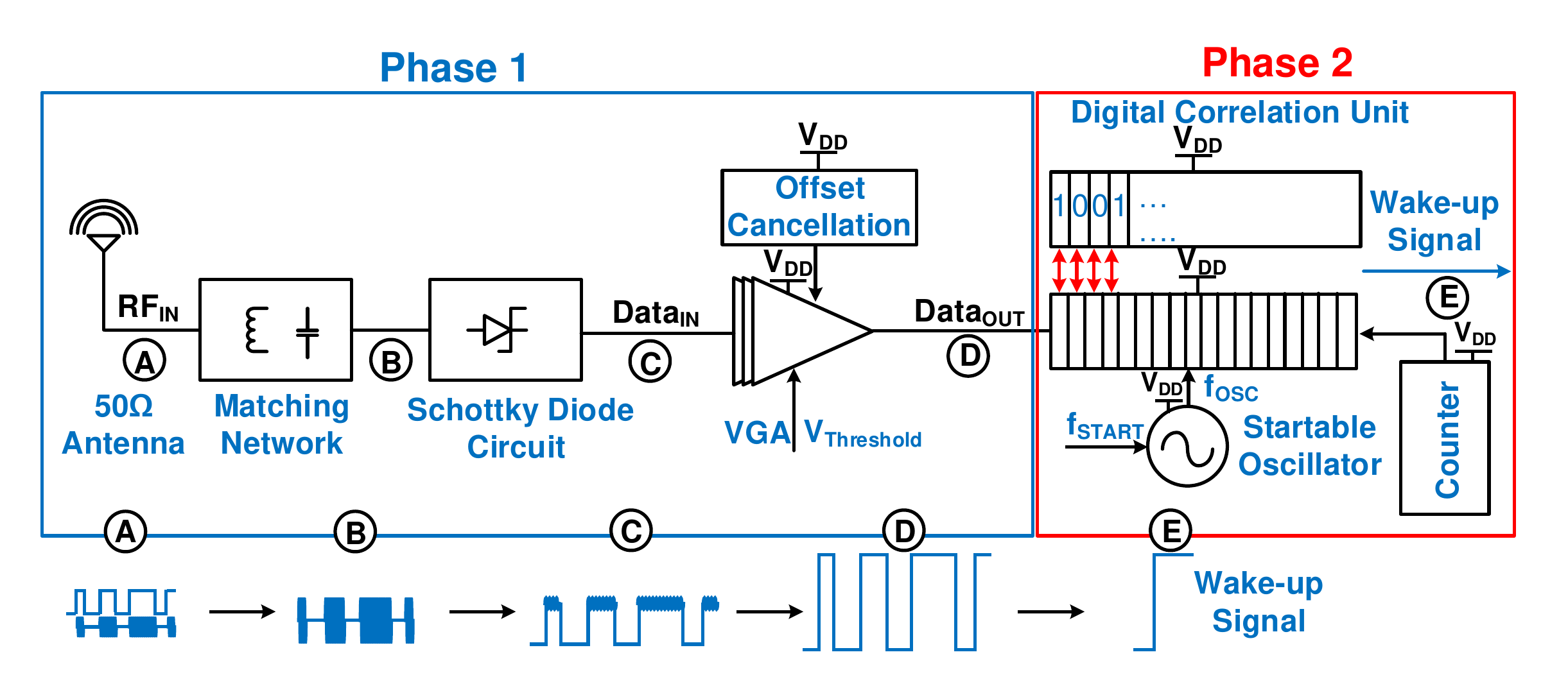}
	\caption{Proposed wake-up receiver architecture}
	\label{fig:2-2}
\end{figure*}
\renewcommand{\b}[1]{\boldsymbol{\mathrm{#1}}}
\section{System Architecture and Analysis}\label{sec:model}

This section describes the architecture of the WuRx. An analytical model for the system is then presented to compare the detection performance to other architectures under varying SNR. The model is also used to evaluate the impact of the length of the digital correlation and the impact on power dissipation. 
\subsection{Wake-Up Receiver (WuRx) Architecture}
The system architecture is shown in Fig. \ref{fig:2-2}. In the first phase, the RF input is filtered by a matching network into a Schottky diode. The diode detects the signal power which is then amplified by a series of amplifiers. The last stage of the first phase is a comparator with an adjustable threshold. Data transitions from the thresholding amplifier determine whether to trigger the activation of the second phase. A startable oscillator is activated with a data transition and digitally correlates the received signal with a pre-determined signature. When a signature has been detected with sufficient accuracy and repetition (both are parameters), the signal is generated to wake-up the primary transceiver of the IoT system.
\par
The first stage is always operating and can be tuned to different center frequencies depending on the front-end filter. Our system is designed for a center frequency of 750MHz. Details of the circuits are described in Section \ref{sec:sysimp}. The power cost of our implementation of the first stage and its sensitivity is 1.48 $\mu$W and -50dBm for a data rate of 200kbps. The second stage has an energy cost of 0.2$\mu$W during the time it is awakened and runs a correlation. These parameters are used in the analytical model of the system presented in the next subsection.
\par
The Schottky diode has a similar if not better conversion gain as the subthreshold slope of a CMOS transistor. For the bandwidth needed for high input frequency and data rate, the diode is biased above zero. The proposed CMOS Schottky diode provides a conversion gain of 660 at the required input frequency of 750MHz when biased at 50mV and 1.25$\mu$A. Whereas, the CMOS ED would require Vth $\sim$ 250mV with the same current biasing to achieve equivalent data rate and the same bandwidth which results in higher power consumption.
\subsection{System Model}
An analytical model of the receiver is used to compare different architectures. Assuming two hypothesis of whether a signature $\b{s}$ is transmitted or not:
\begin{itemize}
	\item $H_{0}$:  The sequence $\b{s}$ was not transmitted.
	\begin{itemize}
		\item $H_{0A}$: No signal was transmitted
		\item $H_{0B}$: A  sequence $\b{s}'\neq \b{s}$ was transmitted.
	\end{itemize}
	\item $H_{1}$: The sequence $\b{s}$ was  transmitted .
\end{itemize}
The probability of false alarm is defined as $P_{FA} = Pr(\text{declare }H_1|H_0) $, and the probability of detection is defined as  $P_{D} = Pr(\text{declare }H_1|H_1)$.
\par
For our analysis, we define the transmitted signal, $\boldsymbol{x}$, as 
\begin{equation}
\b{x}=\begin{cases}
\b{0} & H_{0A}\\
\b{s}' & H_{0B}\\
\b{s} & H_{1}
\end{cases}
\end{equation}
Since we are using OOK modulation,   $\b{s}$ is a binary sequence  of length $L$ having $d$ ones.  The WuRx is triggered when either $\b{s}$ or $\b{s}'$ has the first bit equal to one.  The  received signal is modeled as $z =x + n$
where $\b{n}$ is the additive white Gaussian noise (AWGN) vector with zero mean and variance $\sigma^{2}$ and the signal-to-noise-ratio (SNR) is equal to $1/\sigma^{2}$ for a normalized input signal.
\par
By using this signal model and the possible events, we model three different WuRx architectures for comparison: one with only an energy detector (ED), one with two-phase wake-up by adding a correlator (Corr), and one with a matched filter (MF).
\paragraph{Energy Detector WuRx (ED)}
The decision made by an energy detector can be modeled as follows:
\begin{equation}
\text{ED declares}\begin{cases}
H_{0} & z[1]\leq\lambda\\
H_{1} & z[1]\geq\lambda
\end{cases}
\end{equation}
where $\lambda$ is a threshold and $z[1]$ is the first element of the vector $\b{z}$. Due to the simplicity of the energy detector, it is unable to differentiate between $H_{0A}$
and $H_{0B}$. $P_{FA}$ due to ED is formulated in Appendix-A.

\paragraph{Two-Phase WuRx with Correlator (Corr)}
Instead of waking up the receiver for every signal that crosses the energy threshold, a correlator improves detection by comparing the received signature to the sequence $\b{s}$.
In the first phase, when the first element of the received vector $z[1]$  crosses the threshold, of $\lambda$, the correlator of the second phase is activated.  The binary vector $\bar{\b{z}}$ is obtained using the ED threshold $\lambda$ and is compared to the sequence $\b{s}$. The system declares $H_{1}$ if the received sequence $\bar{\b{z}}$ differs by at most $l$ bits from the reference sequence $\b{s}$. This can be expressed as follows
\begin{equation}
\bar{z}[k]=\begin{cases}
0 & z[k]\leq\lambda\\
1 & z[k]>\lambda
\end{cases}
\end{equation}
\begin{equation}
\text{WuRx declares}\begin{cases}
H_{0A} & \bar{z}[1]=0\\
H_{0B} & \bar{z}[1]=1,\ \sum_{i=2}^{L}\bar{z}[i]\oplus s[i]>l\\
H_{1} & \bar{z}[1]=1,\ \sum_{i=2}^{L}\bar{z}[i]\oplus s[i]\leq l
\end{cases}
\end{equation}
where $\oplus$ is the binary xor operator.
Since the correlator uses the entire sequence for comparison, it is expected to have a probability
of false alarm lower than the ED. However, at low SNR, if we consider the exact match only ($l=0$), any mistakes in the bits would lead to a missed detection. The derivation of $P_{FA}$ and $P_{D}$ for this architecture is provided in Appendix-B.

\paragraph{OOK Matched Filter  WuRx (OOK MF)}
A third architectural option is to use a matched filter to compare  the analog values of
$\b{z}$ with the signature $\b{s}$. By using the energy of the entire received signal prior to using a symbol-by-symbol threshold, one can expect an improved effective SNR. The improvement comes at the cost of the power from an analog correlator. 
\par
The MF operation can be written as $\text{\ensuremath{\eta=\b{z}^{T}\b{s}}}$
\begin{equation}
\text{MF declares}\begin{cases}
H_{0} & \eta\leq\lambda\\
H_{1} & \eta>\lambda
\end{cases}
\end{equation}
 Since,
$\boldsymbol{s}$ is a binary sequence, $\eta$ can be rewritten as
the sum of the elements of $\b{z}$ at the location where
$s[i]$ is equal one
\begin{equation}
\eta=\sum_{\{i:s[i]=1\}}z[i]
\end{equation}
Since $\eta$ is the sum of multiple values of $\b{z}$,
it is expected to be more robust to false alarms due to noise specially at low SNRs compared to ED, which
only considers the first bit of $\b{z}$. However, the
MF does not take into consideration the values of $\b{z}$
where $\b{s}$ is equal to zero. Hence, many different values
of $\b{z}$, which might correspond to wrong signatures $\b{s}'$, will map to the same value of $\eta$. This  would lead to more false alarms  compared to the correlator at high SNRs. Further trade-off analysis of OOK MF across different parameters is provided in Appendix-C.

\paragraph{BPSK Matched Filter  WuRx (BPSK MF)}
To avoid the shortcoming of an OOK MF where multiple values map to the same $\eta$, a BPSK modulated sequence needs to be used such that $\b{s}$ is the sequence of $\pm1$ instead of zeros and ones. The statistic $\eta$ in that case only attains the maximum value if the received vector $\b{z}$ matches signature $\b{s}$.  Although the BPSK MF performs better, it needs a mixer instead of a square-law device to downconvert the signal, leading to higher power consumption.
The analysis of BPSK MF across different parameters is provided in Appendix-D.

\subsection{Comparison of Architectures}
\label{subsec:comp_approach}
\begin{figure}[t!]
	\centering
	\subfloat[SNR = 6 \label{fig:roc_6}]{\includegraphics{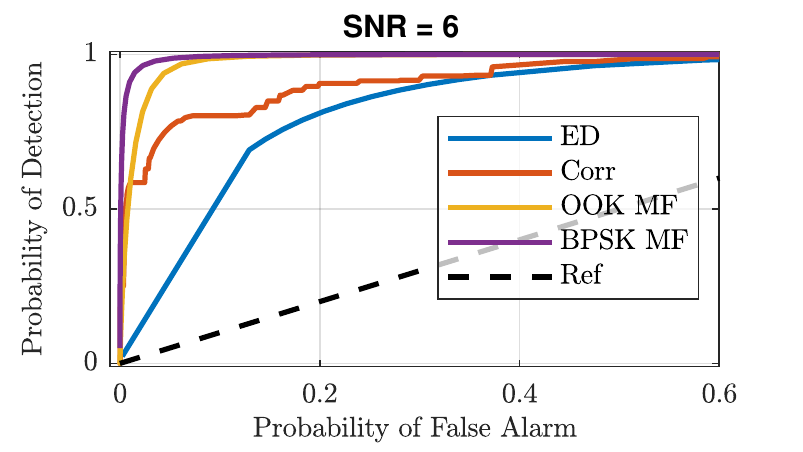}} \\
	\subfloat[SNR = 15 \label{fig:roc_15}]{\includegraphics{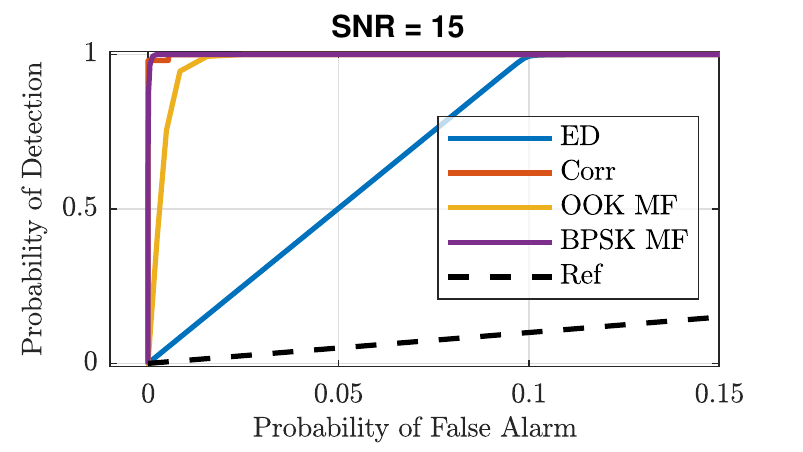}} 
	\caption{ROC curves of ED, MF, and Corr at (a) SNR of 6dB and (b) SNR of 15dB.}
	\label{fig:roc}
\end{figure}
This section compares the performance of the above architectures.  Since, there is a trade-off between the probability of false alarm and the probability of detection, we use the receiver operation characteristics curves (ROC) to show this trade-off~\cite{detection_textbook}. In the following, both Corr, and
MF use a sequence length $L=8$  and $d=4$ ones. The ROC curve is obtained
for ED and MF by sweeping the values of $\lambda$ in the range $[-2,2]$ and $[-1.25L,1.25L]$
respectively. The BPSK MF signal power is normalized for a fair comparison. For the correlator, $l$ is swept from 0 to $L$
and for each value of $l$, the threshold is swept between $[-2,2]$.
The expressions used to generate these curves are from the Appendices. 
The prior probabilities are chosen as  $P(H_{0A})=0.9$, $P(H_{0B})$=0.1$\times$255/256, and $P(H_{1})$=0.1/256. These values are equivalent to 10\% utilization with 256 equally-probable 8-bit sequences.

The ROC curves with SNR of  6 dB and 15 dB are shown in Fig.~\ref{fig:roc}.
At 15-dB SNR, Corr outperforms both MF and ED. Since Corr checks the entire sequence, it achieves
lower $P_{FA}$ compared to ED. OOK MF suffers from being unable to distinguish sequences that 
match $\b{s}$ at its ones while differing at its zeros, and hence,
has a relatively lower $P_{FA}$ than BPSK MF. 
As the SNR drops to 6dB, both MF, by using the energy of the entire sequence, outperform Corr with BPSK MF performing slightly better. Since the correlator relies on binary decisions from the ED, missed detections or errors in the sequence occur more frequently at low SNR.

\subsection{Correlator Design Analysis}\label{subsec:3-3-3}
\begin{figure}[t!]
	\centering
	\includegraphics{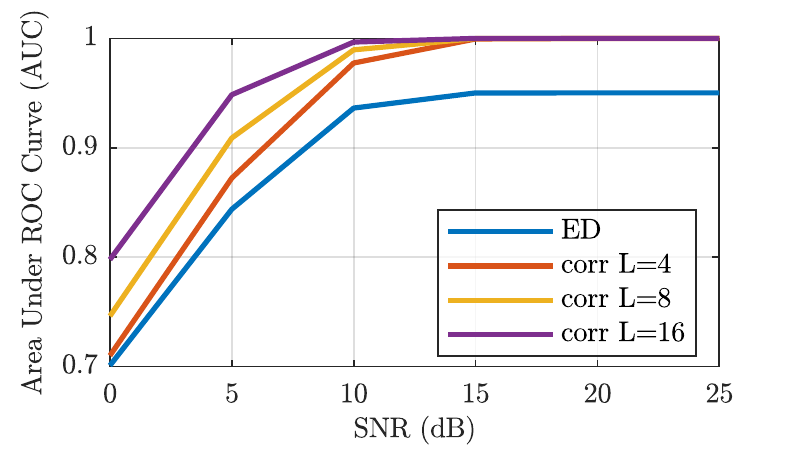}
	\caption{The effect of the correlator length (L) on the WuRx performance measured using  AUC at different SNRs.}
	\label{fig:auc_snr}
\end{figure}

The ROC curves in the previous section show that the values of $P_{FA}$ and $P_{D}$ of Corr are determined by the parameters  $l$, $L$, and $\lambda$. The choice of $L$  is a trade-off between wake-up latency and detection performance as measured by $P_{FA}$ and $P_{D}$.
To determine the effect of our choice of the correlator length $L$ on the performance of the WuRx, we generate a ROC curve for the candidate values of $L$ at different SNRs. The area under the ROC curve (AUC) is used as a comparison metric. An ideal detector would have an AUC value equal to one. The ROC curves are generated using the same parameters as in Section II-B. The results are shown in Fig.~\ref{fig:auc_snr}, from which increasing $L$, improves the performance of the detector but with diminishing improvements at long $L$.  Our WuRx implementation that is described in Section \ref{sec:sysimp} considers the trade-off with latency and chooses $L=8$.

\begin{figure}[t!]
	\centering
	\includegraphics{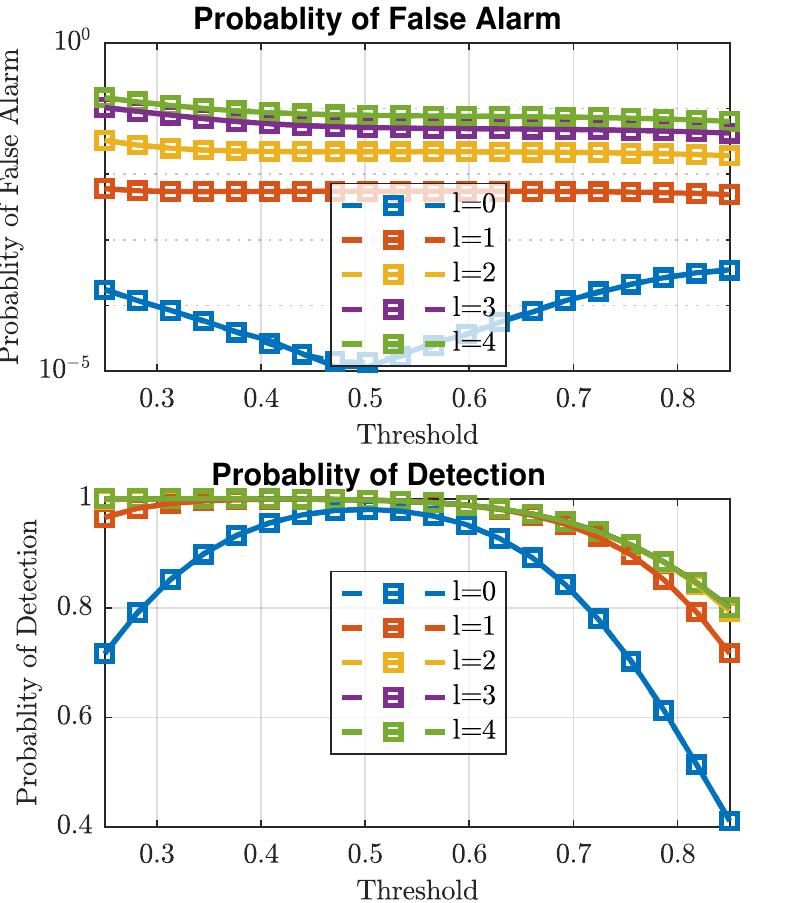}
	\caption{Probability of false alarm and detection as a function of the threshold of an 8 bit correlator for different number of bits, $l$, allowed to be wrong while still declaring a match. Both simulation results (solid lines) and analytical results (dashed lines with markers) overlap. }
	\label{fig:l_thresh}
\end{figure}

With $L=8$, the parameters $l$ and $\lambda$ can be chosen to optimize system performance.  The effect of changing the threshold for different $l$ at SNR=15dB  is shown in Fig.~\ref{fig:l_thresh}. 
The results from  Monte Carlo simulations of $10^6$ sequences for each hypothesis and the derived analytical expressions derived in Appendix-B agree and the lines overlap. 
As $l$ increases, the probability of detection increases but at the cost of substantially increasing (100x) $P_{FA}$. To avoid erroneous wake-ups of the receiver, our implementation uses  $l=0$. At lower SNR of 10dB, the PFA shows 10x increase leading to increase of the overall receiver energy cost. At the same time, the probability of detection reduces and falls below our target design of PD$>$0.99. For L=16, we can see a 10x improvement for the probability of false alarm (PFA) which is important design aspect as it sets the average power consumption. However, the latency will be double in that case.

\begin{figure}[t!]
	\centering
	\includegraphics{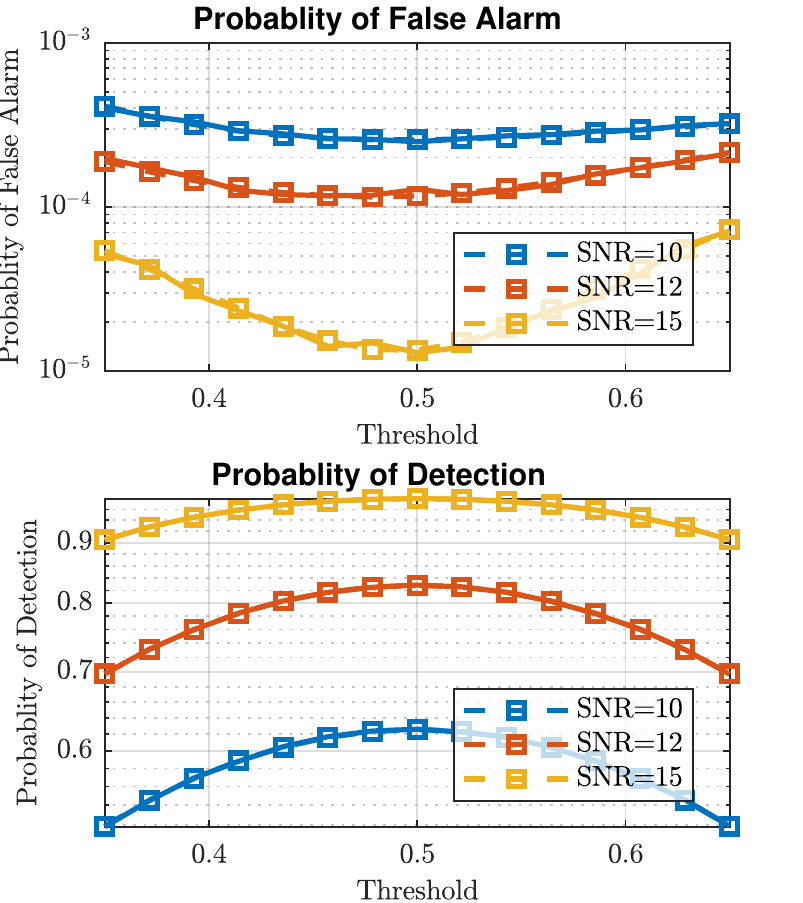}
	\caption{Probability of false alarm and detection as a function of the threshold of an 8 bit correlator for different SNRs. Both simulation results (solid lines) and analytical results (dashed lines with markers) overlap.}
	\label{fig:snr_thresh}
\end{figure}

The effect of varying the threshold on the system performance at different SNRs is shown in Fig.~\ref{fig:snr_thresh}.  In this model, setting the threshold to 0.5 yields the best detection performance for any SNR. This result matches intuition since the noise distribution is gaussian and the sequences are binary. For the implementation in the next section, the realistic dependence of the threshold on offset and signal energy is discussed. Since noise leads to misinterpreting the received signal in $H_{0}$ as $\b{s}$, the lower the SNR the higher the probability of false alarm.

\subsection{Energy Analysis}
The expected energy consumption $E$ of any receiver is related to the probability of false alarm and detection  using
\begin{equation}
E=E_{\text{WuRx}}+(P_{D} P(H_1) +P_{FA} P(H_0))E_{\text{Rx}}
\end{equation}
where $E_{\text{WuRx}}$ is the energy of the WuRx and $E_{\text{Rx}}$
is the energy of the primary data transceiver of the leaf node. 
Since,  $E_{\text{Rx}}>>E_{\text{WuRx}}$, we want
a WuRx with  the lowest $P_{FA}$ for a given $P_{D}$. 

We optimize $E$  in our two-phase implementation by reducing both $E_{\text{WuRx}}$ and $P_{FA}$.  $E_{\text{WuRx}}$ can be expressed as
\begin{equation}
E_{\text{WuRx}} = E^{ED} + (P(H_0) P_{FA}^{ED} + P(H_1) P_{D}^{ED} ) E_{Corr}
\end{equation}
$P_{FA}$ is the result of activating both stages under $H_0$ and can be expressed as
\begin{equation}
P_{FA} = P(H_0) P_{FA}^{ED}  P_{FA}^{Corr} 
\end{equation}

To compare the single-phase and two-phase system, we set a target value of $P_{D}$ and optimize both systems for minimal energy while meeting the required $P_D$. For ED, we optimize the threshold and for Corr, we optimize the threshold and $l$.  The value of the thresholds considered is within 0 and 1 with a step of 0.1 for both systems and $l$ is from 0 to $L$. The optimal parameters are found by trying all possible combinations.

In this analysis, the minimal $P_D$ given by $\gamma$ needs 
to be realized with at most $q$ re-transmissions, and hence, the probability of at least one transmission out of $q$ being detected should be greater than $\gamma$.
\begin{equation}
P(\text{det. at least one of \ensuremath{q}})=1-(1-P_{d})^{q}\geq\gamma
\end{equation}
which is equivalent to 
$
P_{D}\geq1-(1-\gamma)^{1/q}
$.

\begin{figure}[t!]
	\centering
	\includegraphics{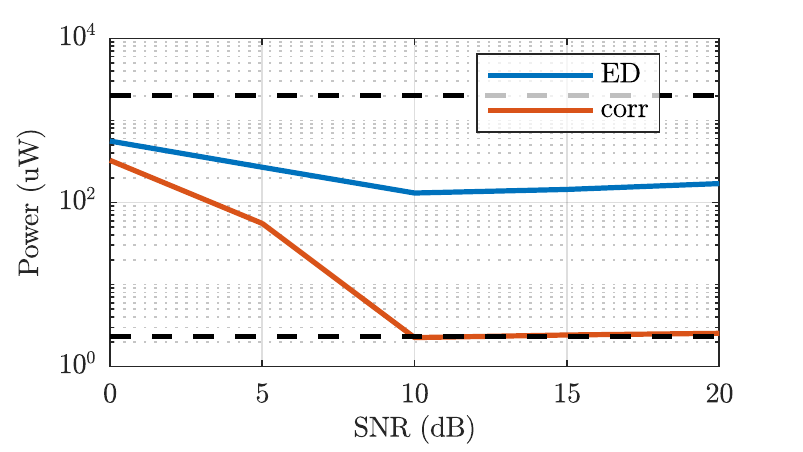}
	\caption{The expected energy consumption of ED and Corr when parameters are optimized as a function of SNR.}
	\label{fig:energy}
\end{figure}

 We consider $\gamma=0.99$ and $q=5$ making the minimum $P_D$ per transmission equal to $0.6$. The results are shown in Fig.~\ref{fig:energy}. Although the ED circuit by itself has a lower energy consumption than the correlator, it mistakenly awakes the primary transceiver often. This increases the energy consumption of the system by up to $60x$. A correlator, on the other hand at SNR of 10 and above, has $P_{FA}$ approaching zero. Thus, it almost attains the lower bound of power consumption which occurs when receiver $P_{FA}=0$. For SNR$>$10,  the optimal parameters of the correlator are $\lambda=0.5$ and $l=0$, matching the previous results. For SNR$<$10, $l$ has to be increased to meet the required $P_D$ leading to higher $P_{FA}$ and thus higher energy consumption.
\subsection{System analysis summary}
The analysis is intended to show the rationale behind the design choices and the importance of using a correlator as part of the design. We showed the design trade-offs of using matched filter for OOK and BPSK modulation and ED with correlator. We opted to use the correlator as part of a 2 stage design that significantly impact the PFA/PD performance of the overall receiver. We analyzed the impact of number of correlator bits used (L) and the allowance of incorrect bits (l) on both the PFA and PD. In addition, the other novelty revolves around the use of two phase wake up to reduce average power consumption. We proposed the use of CMOS Schottky diode as ED which operates at low bias and high conversion gain compared to CMOS ED. In the second-phase we proposed a data-locked startable oscillator which saves the system significant power consumption using oversampling or DLL compared to prior implementations.

\section{Wake-Up Receiver Circuit Design} \label{sec:sysimp}
This section describes the design and implementation of circuit components in the proposed WuRx system. First, the CMOS integrated Schottky diode as a square-law detector is discussed in detail to achieve the desired performance at a high carrier frequency with low power dissipation. The description of a low-power amplifier with a variable threshold follows. The two elements of the second phase are then discussed: the startable oscillator and the digital correlator. 
\subsection{CMOS Integrated Schottky Diode} \label{subsec:diode}
A Schottky diode is used as an energy-efficient power detector that accommodates the required input frequency and bandwidth. The Schottky diode output voltage can be expressed as $V_{out_1stage}$=$\gamma_{eff}$$\times$$V_{rf}^2$, where ($\gamma_{eff}$) is the conversion gain for the power detector which is dependent on the incident power. 
\par
\begin{figure}[!t]
 	\centering
 	\includegraphics[scale=0.3]{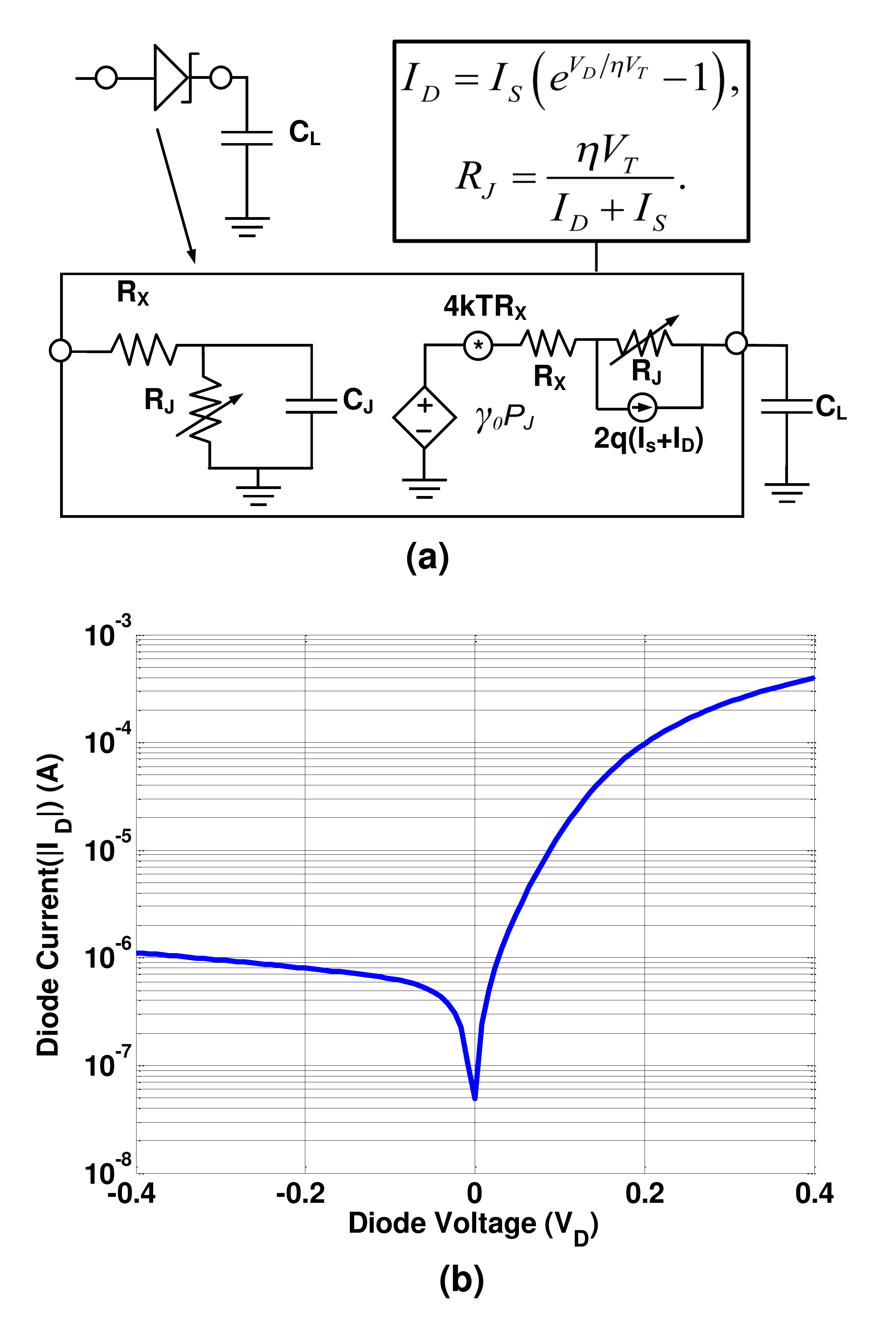}
 	\caption{(a) Schottky diode small-signal model and equations, (b) and measured I-V characteristics for fabricated diode.}
 	\label{fig:2-6}
 \end{figure}

The model of the diode is shown in Fig. \ref{fig:2-6}(a) Where $R_{J}$ is the dynamic equivalent resistance of the diode, $R_{X}$ is a parasitic series resistance, $C_{J}$ is a parasitic capacitance, $P_{J}$ is the equivalent power delivered to $R_{J}$, $\gamma_{0}$ is the DC voltage-sensitivity, and $C_{L}$  represents the load, N is the ideality factor of the diode. The model used for the impedance estimate shown in Table \ref{table:T333}.

\par
Fig. \ref{fig:2-6}(b) shows the measured I-V characteristics of a fabricated diode in a standard CMOS technology. We designed this diode using a Schottky junction formed with a metal contact to the diffusion layer on top of N-well layer similar to the the approach in \cite{1-4-1}. These layers are present in any standard CMOS technology and hence enabling this diode to be usable even if such a device is not explicitly present in the technology library.  
\par
The effective voltage sensitivity of the diode detector as a function of the input frequency $\omega_{rf}$ can be expressed as
\begin{equation} \label{eq:noiseTF1}
\gamma_{eff}=\frac{\gamma_{0}}{1+\omega_{rf}^2C_{J}^2(R_{X}||R_{J})^2}
\end{equation}
\begin{equation} \label{eq:noiseTF2}
\gamma_{eff}=\frac{K}{2(I_{D}+I_{S})}\times\frac{1}{1+\omega_{rf}^2C_{J}^2(R_{X}||R_{J})^2}    V/W
\end{equation} 
$\gamma_{0}$ is a function of the bias point of the diode and increases with smaller bias as $I_{D}$ approaches 0 and bounded by the intrinsic reverse saturation current ($I_{S}$) of the diode, and $K$ is a proportionality constant that depends on the technology. The bandwidth for diode can be expressed as $\omega_{3dB}=G_{md}/C_{J}$ where $G_{md}=I_{D}/(N*V_{T})$. Table \ref{table:T333} lists the parameters for our implementation. A bias  of 1.6$\mu$A corresponding to roughly 50mV is sufficient for 1GHz bandwidth which accommodates our input center frequency of 750 MHz while obtaining a conversion gain, $\gamma_{0}$, of 660.  

\begin{table}[!t]

		\centering
		\setlength{\tabcolsep}{5pt}
	\renewcommand{\arraystretch}{1.5}
\caption{Schottky diode parameters}
	\label{table:T333}
\begin{tabular}{ |c| c|  }
\hline
 Parameter & Value  \\ 
 \hline
 $R_{X}$ & 380$\ohm$  \\ 
 \hline
 $C_{J}$ & 8fF    \\
 \hline
 $I_{S}$ & 0.95$\mu$A \\  
 \hline
 $I_{D}$ & 1.6$\mu$A   \\
 \hline
  N & 1.35   \\
 \hline
\end{tabular}
\end{table}

  \begin{figure}[!t]
 	\centering
 	\includegraphics[scale=0.3]{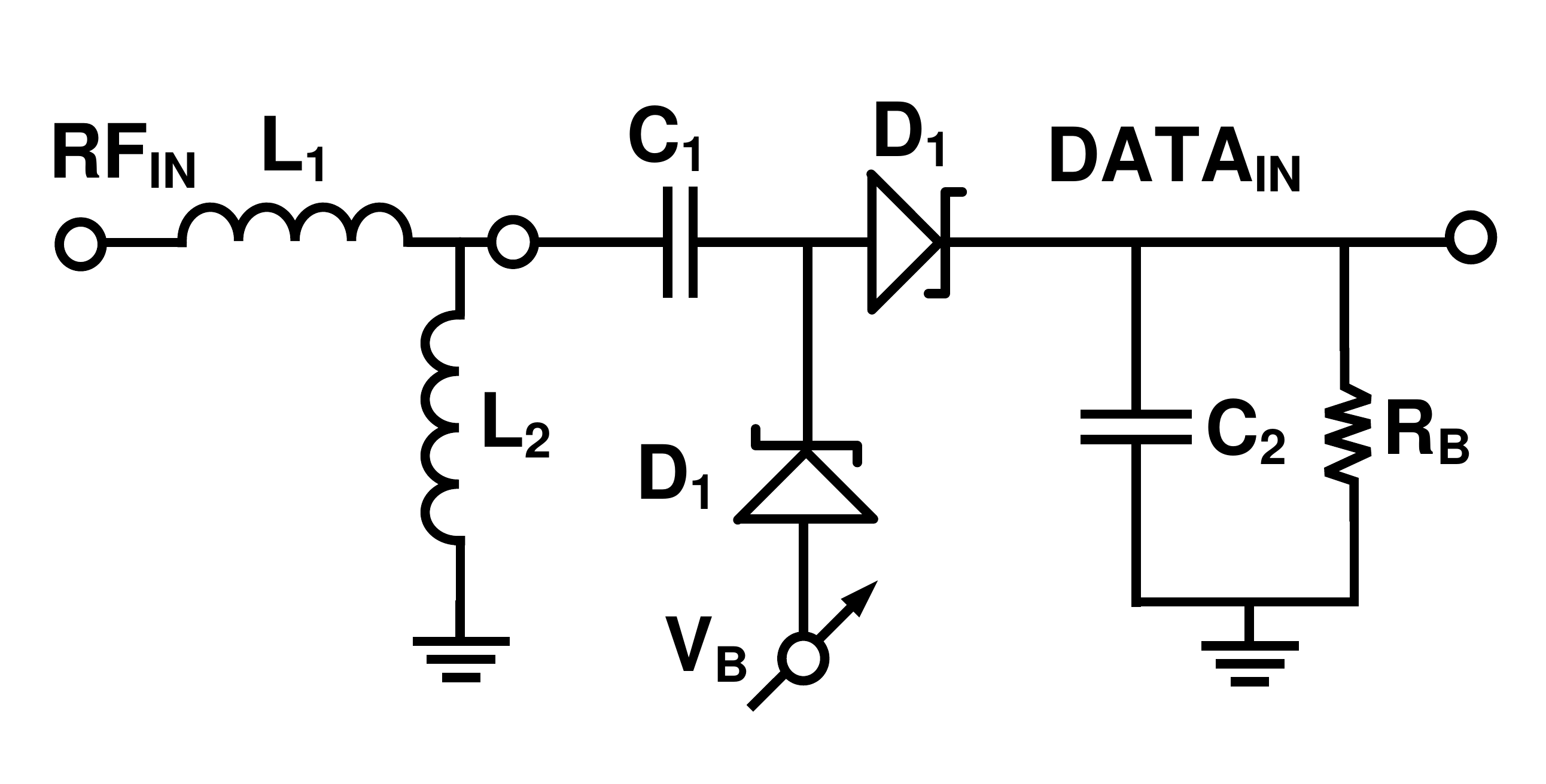}
 	\caption{Energy detector structure using Schottky diode and matching network.}
 	\label{fig:2-33}
 \end{figure}
  
 \begin{figure*}[!t]
	\centering
	\includegraphics[scale=0.6]{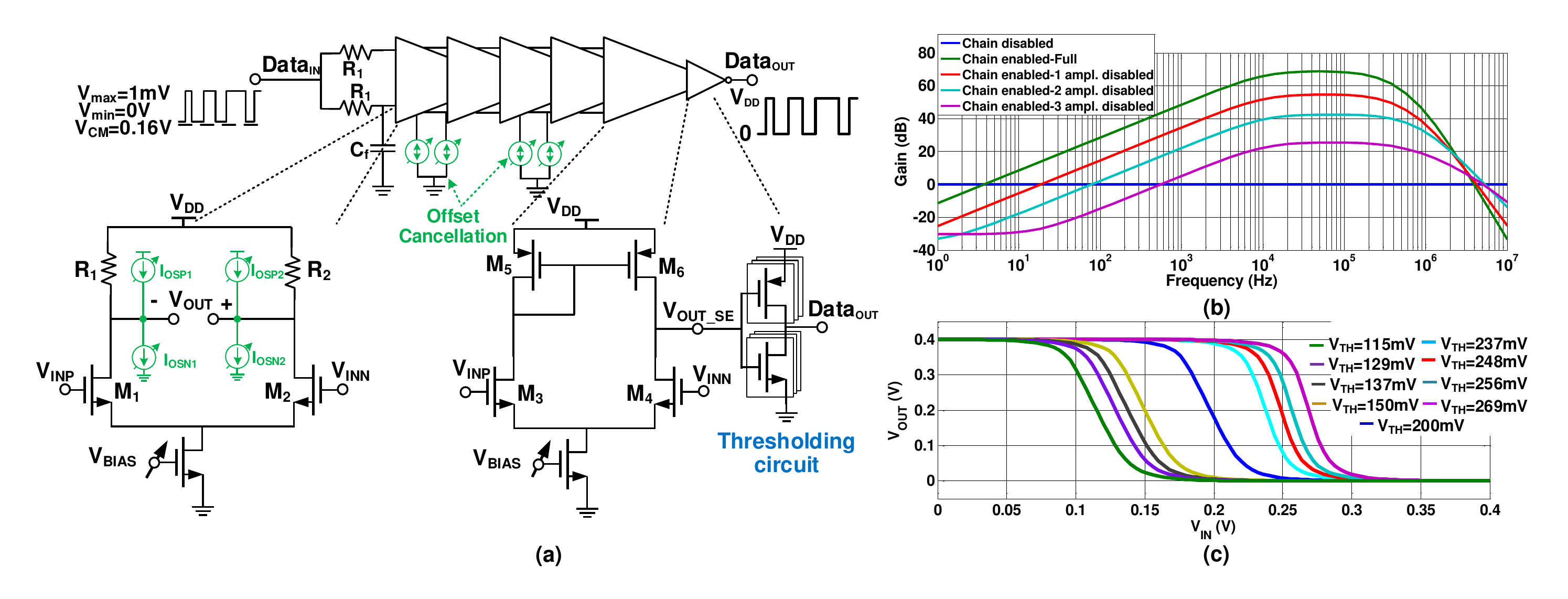}
	\caption{(a) Amplifier chain circuit implementation with offset cancellation, (b) amplifier gain configuration for different input power levels , (c) different threshold voltage configuration.}
	\label{fig:2-7}
\end{figure*}

\par
A Schottky diode doubler stage increases the voltage gain at the cost of increasing the bias voltage. Since the subsequent amplifying stage requires a dc common-mode voltage, increasing the bias voltage does not increase the power cost. In our implementation shown in Fig. \ref{fig:2-33}, we employ a doubler with bias voltage, $V_{B}$=300mV, such that the subsequent amplifier is dc biased at 200mV. As seen at the input of the doubler from the matching network, the input impedance is $R_{IN}$=2.15k$\Omega$ in parallel with $C_{IN}$=3.7pF. The input capacitance seen at the voltage doubler input is mainly parasitic capacitance due to coupling the input from the off-chip matching network. 
\par
The matching network transforms the 50-$\Omega$ impedance of the antenna to the input impedance of the doubler. Due to the finite quality factor (Q$\sim$25) of the off-chip inductors, the achieved gain is 13 dB. The filter is designed to be sufficient for 200-kHz data bandwidth while also providing rejection for interferers. The filter structure is built with $L_{1}$=5.3nH and $L_{2}$=40nH.
 
\par
The equivalent noise sources are also shown in Fig.  \ref{fig:2-6}(a) including thermal noised due to $R_{X}$ and shot noise due to $R_{J}$. The noise power due to shot noise and thermal noise is estimated at 5.12*$10^{-18}$ $A^2$/Hz and 4.4*$10^{-23}$ $A^2$/Hz respectively. As expected for a diode, the shot noise dominates the output noise.
\par
 The noise for the square-law doubler stage can be modeled \cite{1-3-8} as
\begin{equation} \label{eq:nf}
NF_{SD}=1+\frac{N_{o,SD}}{N_{s}G_{P}^2}
\end{equation} 
where $N_{s}$=4kT$R_{s}$ and $N_{o,SD}$ are the thermal noise and shot noise power of the voltage doubler respectively. $G_{P}$ in the expression is the conversion gain of the passive network multiplied by the gain of the voltage doubler. $G_{P}$ depends on the incident power to the power detector which corresponds to lower gain at lower input power leading to higher noise figure. 
\begin{figure*}[!t]
	\centering
	\includegraphics[scale=0.5]{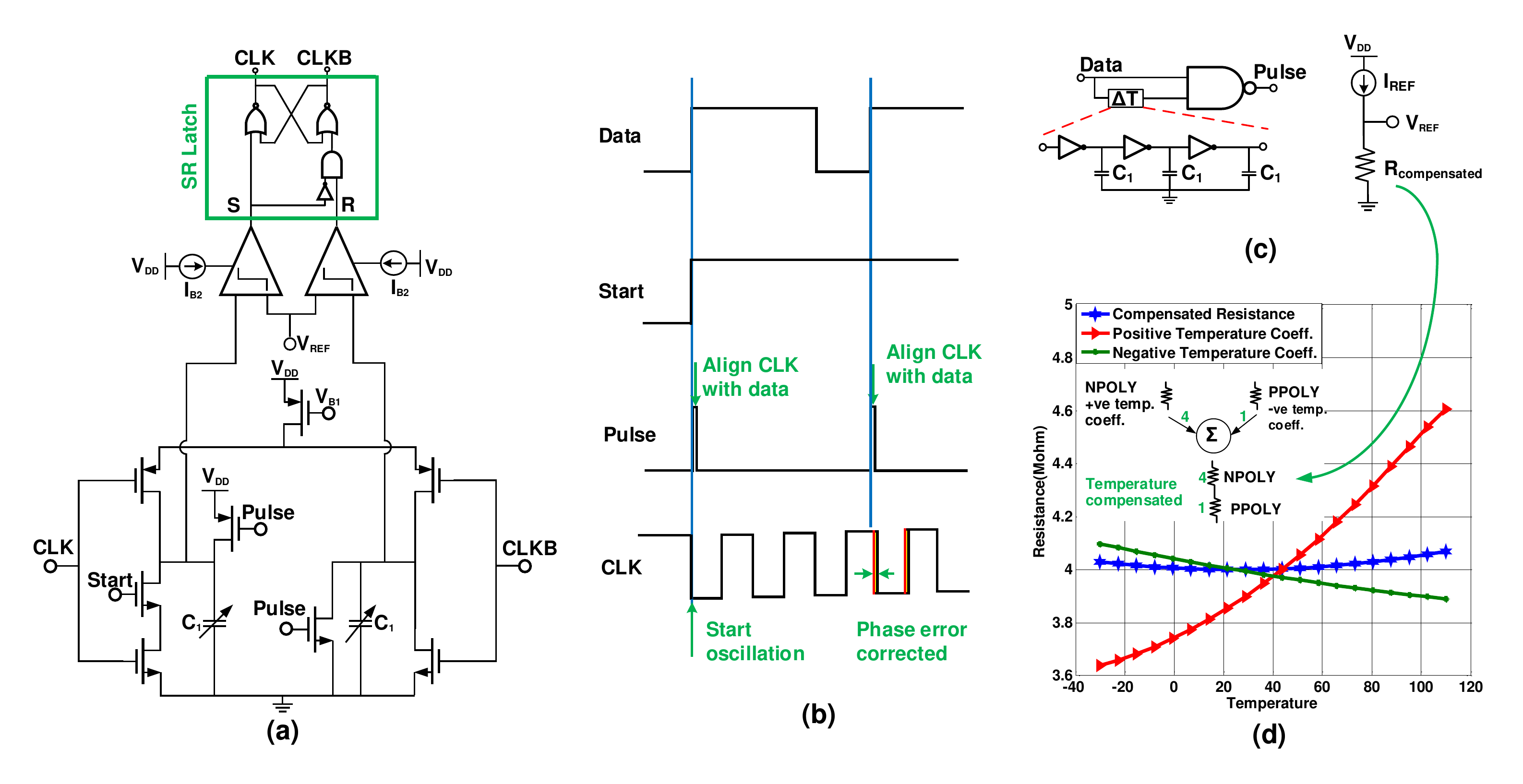}
	\caption{(a) Relaxation oscillator architecture with data locking technique, (b) timing diagram for oscillator operation, (c) pulse generator for data locking technique and reference voltage generator, and (d) zero-order temperature compensation using Npoly and Ppoly.}
	\label{fig:2-9}
\end{figure*}
\par
Multiple doubler stages, as shown in Fig. \ref{fig:2-33}, have been demonstrated \cite{1-3-3} to provide more power gain. In our design, we opt for a single doubler stage due to several considerations. A larger number of stages increases the bias voltage by N while only increasing the power gain by $\sqrt{N}$. A larger diode-capacitor chain impacts data bandwidth and latency. While smaller capacitors or higher bias current can be used, they either results in more noise or more power leading to worse overall performance. 

\subsection{VGA Design}

The down-converted signal at 200kbps is amplified to full scale with the design shown in Fig.\ref{fig:2-7}(a) before being digitally correlated with the signature. The chain consists of five amplifiers followed by a thresholding circuit. The gain control of the amplifier chain is done by bypassing individual stages and disabling unused amplifiers. This approach also saves power dissipation of 200nW per stage. The control signals for bypassing and disabling are determined by detecting the amplitude of the amplified signal at the input of the thresholding circuit. The gain is reduced by turning off each of the first four stages in sequence starting from the last stage. The gain adjustment covers gain range from 66dB to 0dB, as shown in Fig \ref{fig:2-7}(b). 
\par 
The supply voltage of the amplifier chain is 0.4V and the same supply is shared by the digital logic of the second stage. The input common-mode of the first stage from the voltage doubler is 0.2V. Due to the low supply voltage and near-threshold operation of the transistors, the sizing and biasing of the amplifiers are carefully chosen and adaptable. The first four stages are differential amplifiers with resistor loads for good common-mode supply noise rejection \cite{2-1-90}. The last stage uses a current-mirror load for higher gain and single-ended outputs.  The first stages have a larger transistor size and higher quiescent current to improve the overall noise of the chain. The transistors' size for the next three-stage is the same. The input-referred noise of the chain is dominated by the first stage and is 5.1n$V^2$.

 To set the common-mode level of each amplifier in the chain to 200mV, programmability using four digitally-controlled bits is added to the tail-current. The first stage bias can be adjusted by 200nA and the remaining stages by 100nA. The size of each differential pair is chosen so that at its bias current, the $g_{m}/I_{D}$ is maximized for low power operation.  
 \par
 An offset cancellation technique is implemented to reduce the impact of voltage offset at each stage and to ensure the biasing setting of different stages in the chain by having variable current steps to add or subtract current at the output. The offset is calibrated and canceled at the startup of the chip. The calibration is based on monitoring the output voltage where the tail current is varied. The offset is checked to be limited with $<0.5mV$ so that we can cover the range of signal of 1mV which is equivalent to input power of -50dBm. 
\par
The thresholding circuit is an inverter with $\sim$15dB of gain. The PMOS and NMOS networks are implemented as arrays of digitally-selectable devices so that the effective width is programmable. The programmability controls the threshold level for deciding the "0" and "1" value of the received signal as shown in Fig. \ref{fig:2-7}(c). The threshold can be changed from 115mV to 270mV. The threshold is calibrated by applying input and bypassing the amplifier chain. As discussed in Section II, depending on the noise and interference, the variable threshold lowers the probability of false alarm of the primary transceiver as well as control the probability of starting the second phase of the proposed system.

\subsection{Data-locked Startable Oscillator}

To digitally correlate an input data sequence with a signature, the timing of the data must be recovered to sample each data bit.  In this work, the digital correlation is the second phase of the wake-up and does not continuously operate which lowers the average power consumption. While phase-locking approaches such as a delay-locked loop (DLL) or oversampling \cite{1-3-1} can be used to recover the timing, they both require multiple cycles of the clock or a continuously running clock. Our design uses a startable oscillator \cite{5-5} to enable the second phase. These two-approaches will cost at least 50$\%$ more power than the proposed data-locked technique. Also, compared to having single phase architecture (i.e. oscillator always running in the background), the proposed architecture saves 12$\%$ in average power consumption. It is designed for a start-up within a one-bit period and maintains a stable frequency of the target modulation rate of the data across a wide range of temperature variation. The proposed startable oscillator operate at no latency cost, however it has 2dB penalty on the minimum SNR required for data after amplification.
\par
 Fig. \ref{fig:2-9}(a) shows the design of the oscillator. A positive-edge transition from the thresholding circuit creates a pulse that aligns the $start$ signal with the data edge. The timing of the oscillator can be seen in  Fig. \ref{fig:2-9}(b) where the phase difference is corrected at each positive data edge. In this design, we limited the frequency variation across temperature to be $<$5$\%$ using temperature compensation. The frequency can be expressed as $f_{clk}=\frac{I_{B1}}{C_{1}\times V_{REF}}$ and depends on the bias current, $I_{B1}$, and the reference voltage, $V_{REF}$. 
 \par
 As shown in Fig. \ref{fig:2-9}(d), the compensation is implemented by using a resistor as the reference to generate  $V_{REF}$ and $I_{B1}$ is a fixed ratio of $I_{REF}$. The resistor comprises a positive temperature coefficient resistor (Npoly) in series with a negative temperature coefficient resistor (Ppoly) with the sizing ratio of 4:1. The resulting resistance varies within $<$0.1$\%$ across a temperature range of [-30 to 110]. As a result, the variation of the free-running frequency of the oscillator reduces from $\pm$15$\%$ to $\pm$5$\%$ which is composed of resistor variation and reference current variation. Toggling and synchronizing the startable oscillator with data transitions using a sequence with an 8-bit run length is sufficient to lock the oscillator frequency to within a small deviation ($<$1$\%$) from the input frequency. The proposed techniques do not cause any power penalty and can be used for low power applications to avoid the use of expensive crystal oscillators or any external calibration techniques.
\par 
Another important feature in the proposed oscillator is a fast start-up. The set dominant SR latch avoids having the "CLK" and "CLKB" signal both at the forbidden state, hence avoid the need for any start-up circuits that requires multiple cycles to startup. It starts up with the correct oscillation frequency at the first cycle since it only waits for the "start" signal to oscillate with the previously set bias current. That start signal is asserted when the incident data changes from '0' to '1'. This represents the start of the packet that carries the signature to wake-up the primary transceiver. The proposed techniques are sufficient to correctly sample the incident packet of 40 bits and achieve BER of less than 10$^{-3}$. The proposed oscillator consumes a power of less than 200nW at 200kHz oscillation frequency.

\subsection{Digital Correlator}
As shown in Section \ref{subsec:3-3-3}, since interferers are present in the unlicensed frequency band around 750MHz, a false alarm for the primary transceiver is possible when using only an energy detector leading to a large power penalty. After a '0' to '1' transition is detected from the thresholding circuit of the energy detector, a digital correlator matches the preset signature with the received data hence minimizing the probability of false wake-up. The first step to check for a preamble sequence.  A 4-bit counter activates along with an 8-bit shift register which compares the received data to an externally-defined preamble sequence within 16 clock cycles. If the preamble is not found, the counter disables the oscillator to save power until another transition is detected. The preamble signature, when detected, enables the second step in detection. A 5-bit counter and a 32-bit shift register are activated to store the data after the preamble. The 32-bit data provides information for the selected channel and the signature from the transmitter. The signature is compared with the preset signature and, if a match is found, the wake-up signal is asserted. At the end of cycling through the 5-bit counter, the startable oscillator is disabled. This entire second phase of detection after cycling through the two steps stays off until another '0' to '1' data transition occurs. The correlator is used to improve the PFA as discussed in Section \ref{subsec:3-3-3}. In our test chip, this digital correlator is also used to check the BER of the received data stream.

\section{Measurement Results}\label{sec:MeasRes}

\begin{figure}[!t]
	\centering
	\includegraphics[scale=0.3]{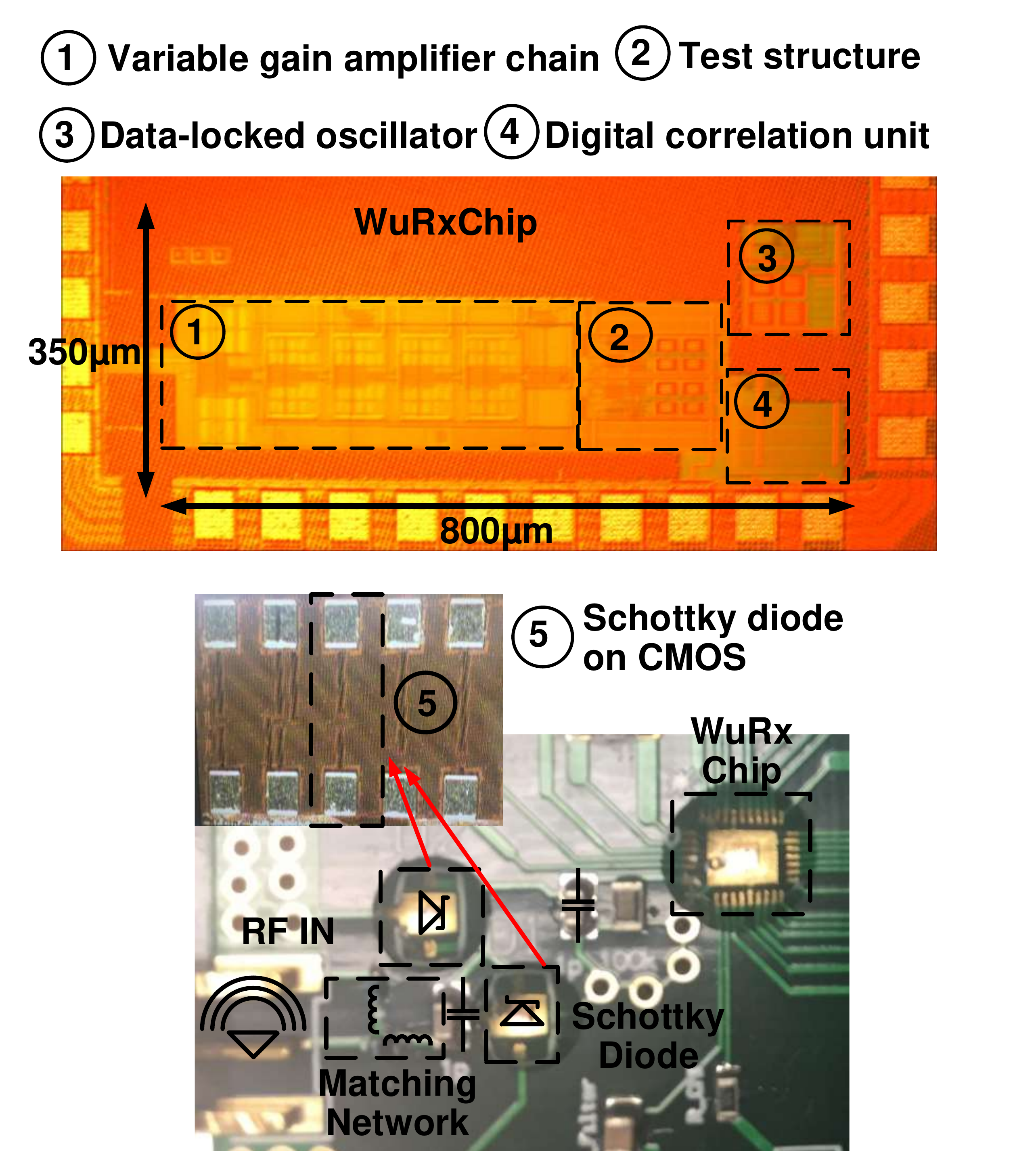}
	\caption{Die micrographs of the WuRx chip and Schottky diode, and PCB photograph of the 
		overall system.}
	\label{fig:2-12}
\end{figure}

The proposed WuRx is implemented in 65-nm CMOS technology and occupies an active area of 800$\mu$m$\times$350$\mu$m as shown in Fig. \ref{fig:2-12}. The implemented chip consists of the baseband processing and amplification,  digital correlator, and generation of the wake-up signal. The CMOS Schottky diode is part of a second die and is connected by wire-bonding between the dice. Due to the low bandwidth at baseband, this assembly does not impact the performance. The input matching network is connected to the diodes by wire-bonding. The entire assembly is shown in Fig. \ref{fig:2-12}.
\par
\begin{figure}[!t]
	\centering
	\includegraphics[scale=0.33]{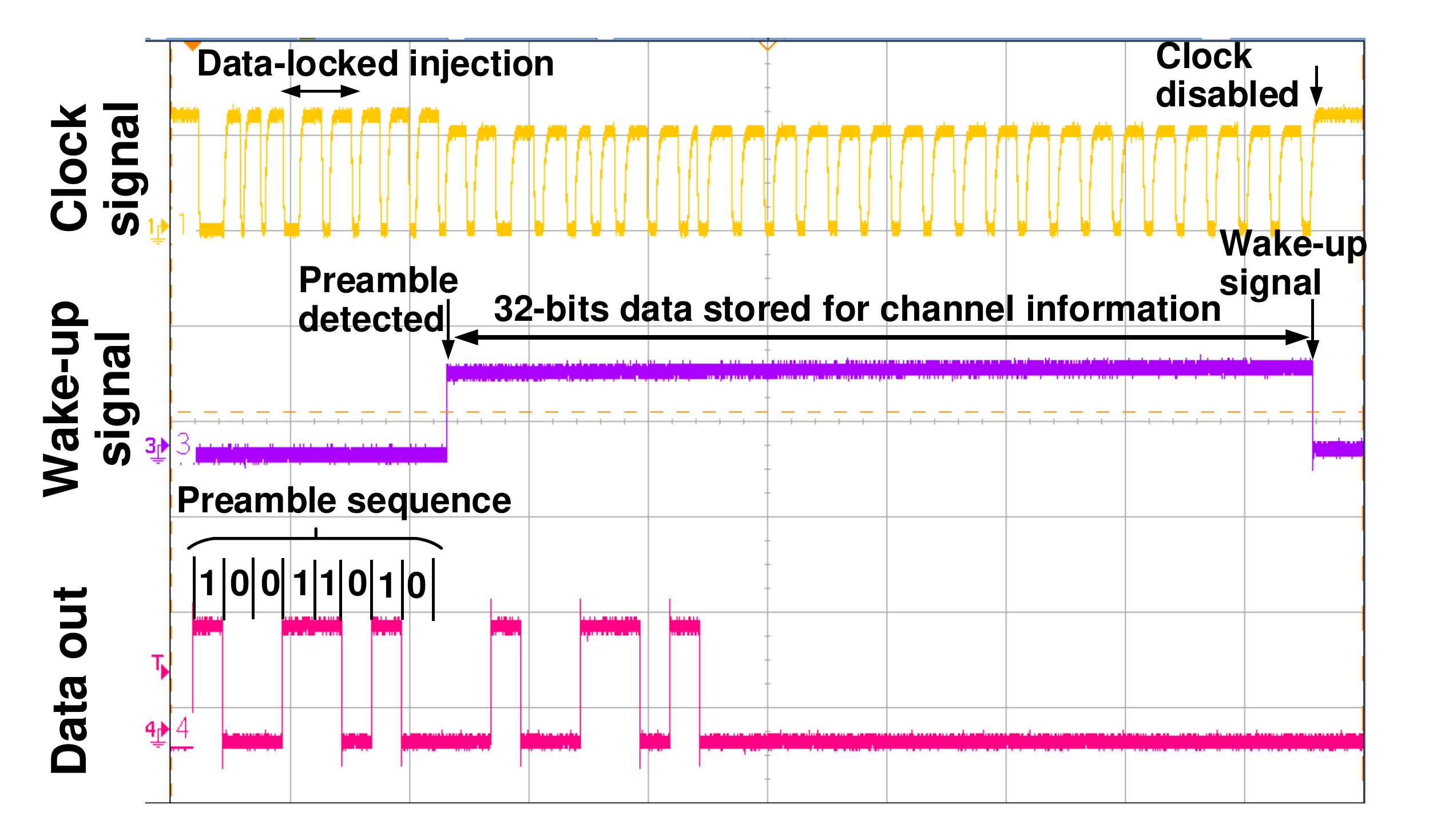}
	\caption{ Measured  detection of the preamble by the WuRx and generation of the wake-up 
		signal with $P_{IN}$=-50dBm and data rate of 200kbps.}
	\label{fig:2-13}
\end{figure}

Fig. \ref{fig:2-13} shows the received data sequence matching the target preamble sequence on an oscilloscope. The shift-register data setting the preamble sequence of "10011010" along with other control bits are loaded using a pattern generator and data acquisition unit (NI PXIe-1082). The same pattern generator produces the OOK pulse modulation that is externally driven to the frequency synthesizer. The clock of the startable oscillator starts when the data changes from 0 to $V_{DD}$ and searches for the preamble sequence within 16 clock cycles. The measurement also shows the oscillator correcting the clock phase based on the data transitions. The wake-up signal is generated to initialize the primary transceiver after storing 32-bits of channel information. After this, the data-locked oscillator is disabled to save power.
\par
\begin{figure}[!t]
	\centering
	\includegraphics[scale=0.22]{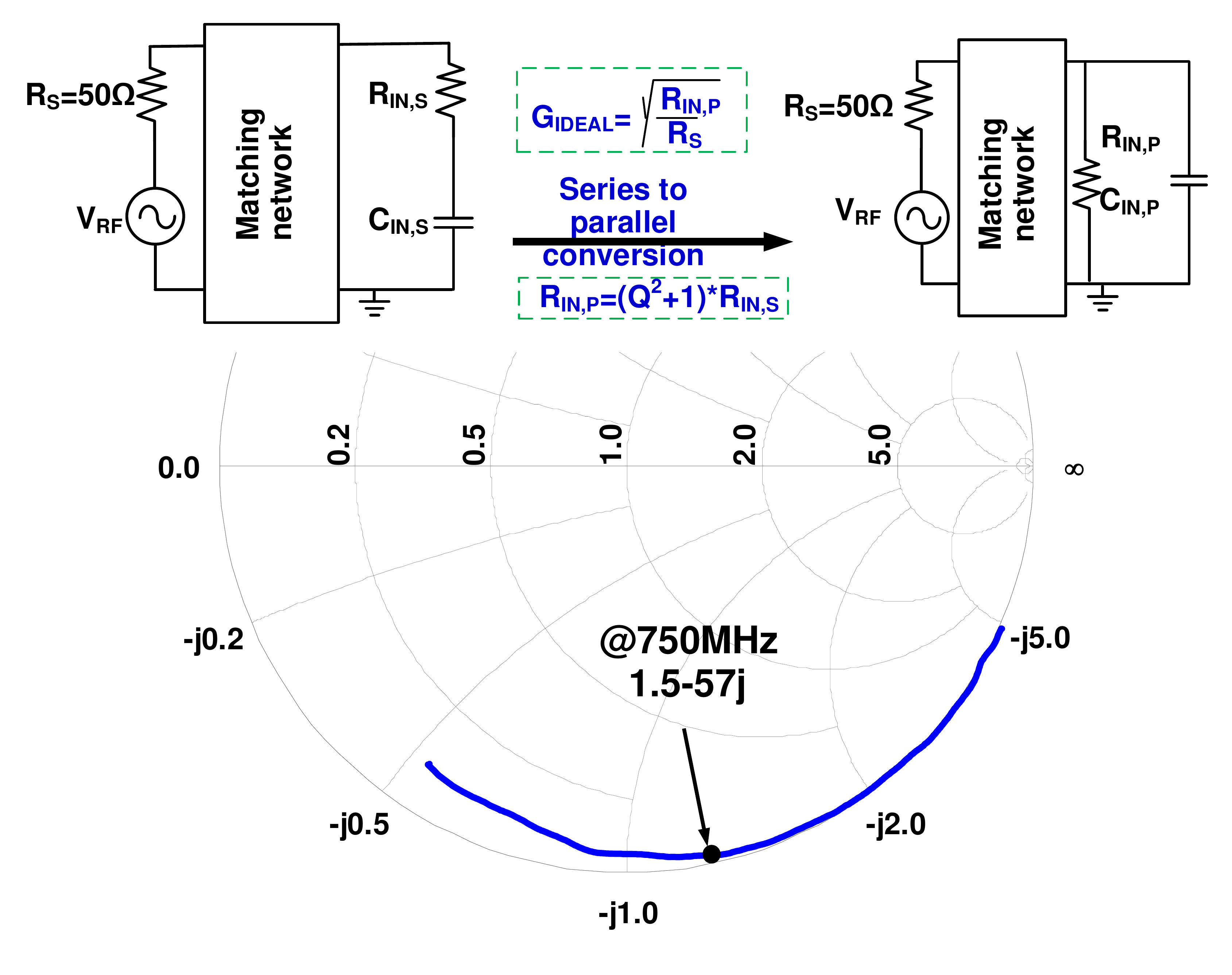}
	\caption{Measured input impedance of the Schottky diode doubler structure at the operating biasing conditions.}
	\label{fig:2-4}
\end{figure}

  
The fabricated Schottky diode input impedance is first characterized using a vector network analyzer (VNA HP8720D) by sweeping from 100MHz-1.8GHz as shown in Fig. \ref{fig:2-4}. At the desired carrier frequency and bias condition of the doubler structure, an input impedance of (1.5-57j) is measured.

\par
\begin{figure}[!t]
	\centering
	\includegraphics[scale=0.31]{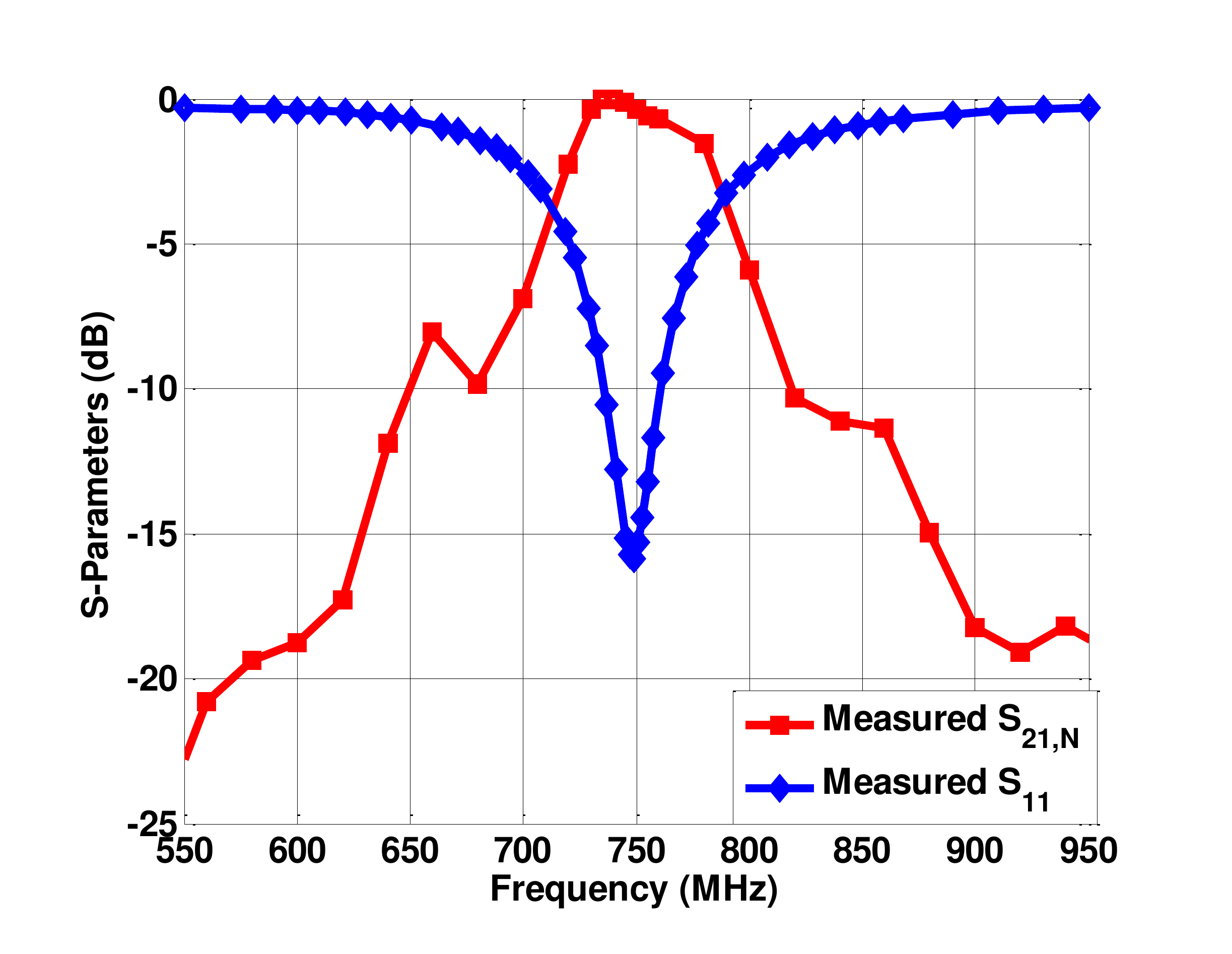}
	\caption{$S_{11}$ and normalized $S_{21}$ measurements. The matching network provides a passive gain of 13dB.}
	\label{fig:2-14}
\end{figure}

The LC matching network is built on a PCB to achieve 50-ohm matching for the diode detector and bonding parasitics, and provide passive gain. Fig. \ref{fig:2-14} shows the measured input matching $S_{11}$ from 500 MHz to 1 GHz. The measured and normalized $S_{21}$ shows the band-pass filtering of the matching network around the carrier frequency. The bandwidth is sufficient for the 200kbps data rate and reduces the integrated in-band noise. The simulated passive gain of the matching network shows roughly 13dB of gain at the carrier frequency.


\begin{figure}[!t]
	\centering
	\includegraphics[scale=0.27]{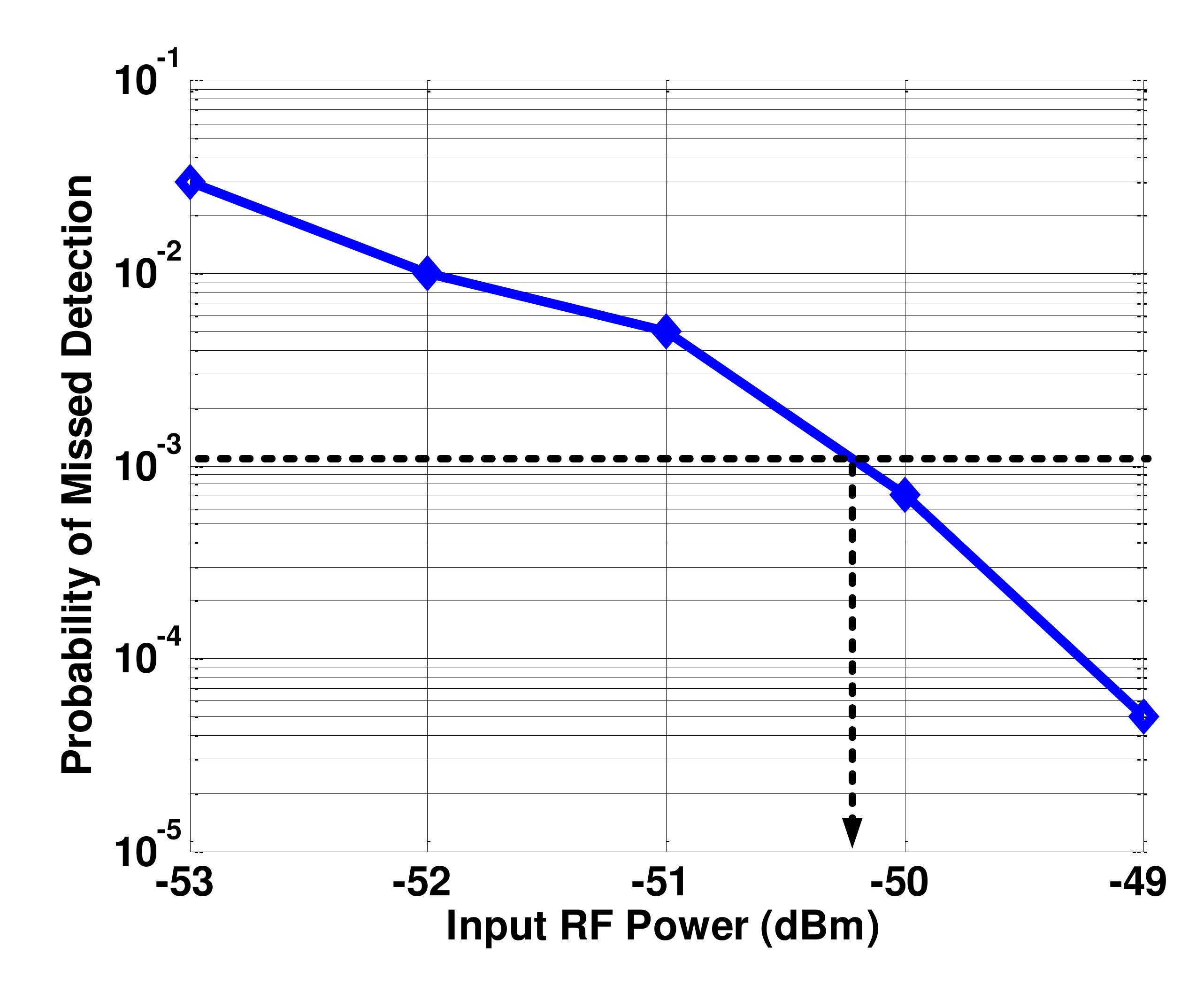}
	\caption{Measured  missed  detection  versus input power of the WuRx.}
	\label{fig:2-15}
\end{figure}
\begin{figure}[!t]
	\centering
	\includegraphics[scale=0.24]{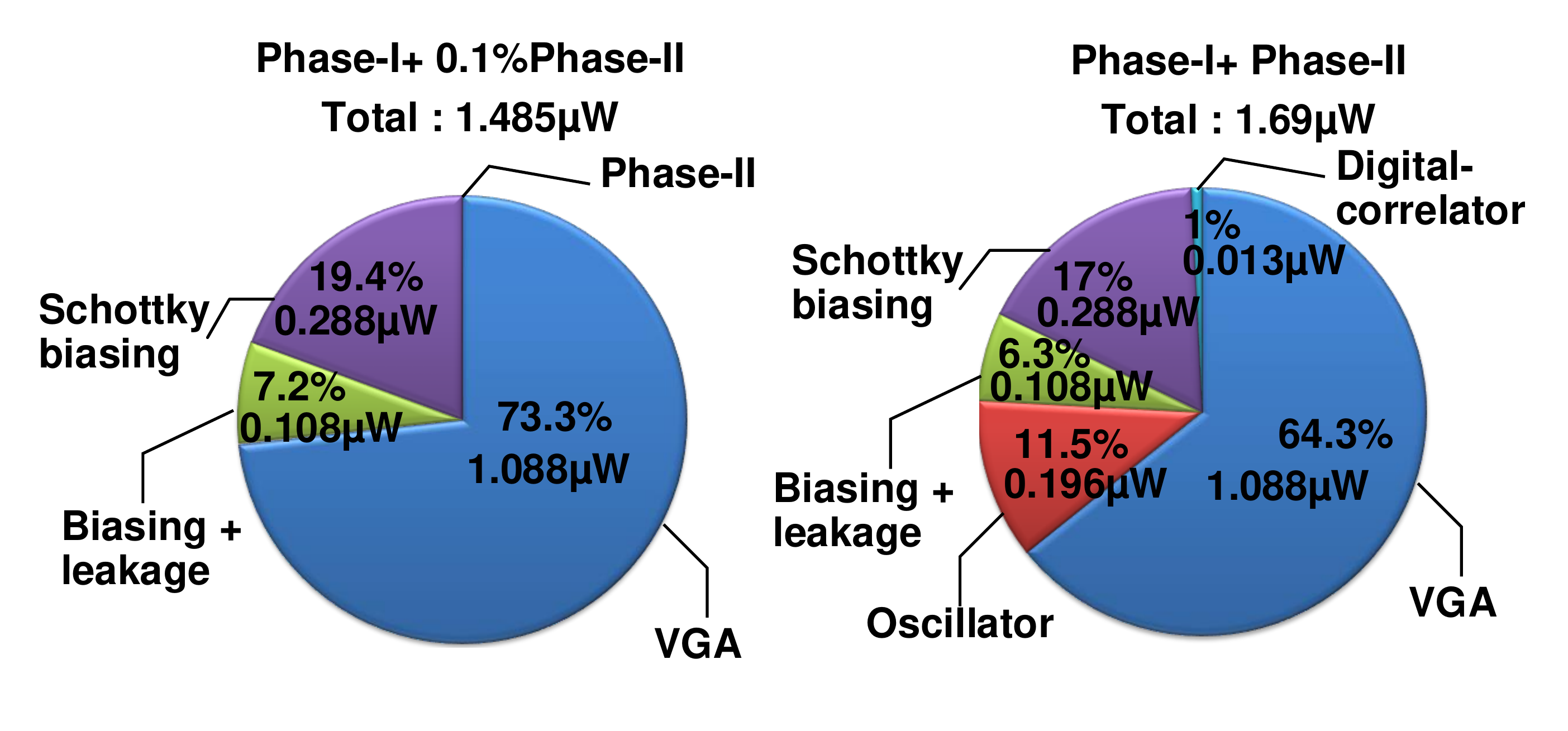}
	\caption{Power breakdown of proposed WuRx.}
	\label{fig:2-16}
\end{figure}

\begin{figure}[!t]
	\centering
	\includegraphics[scale=0.24]{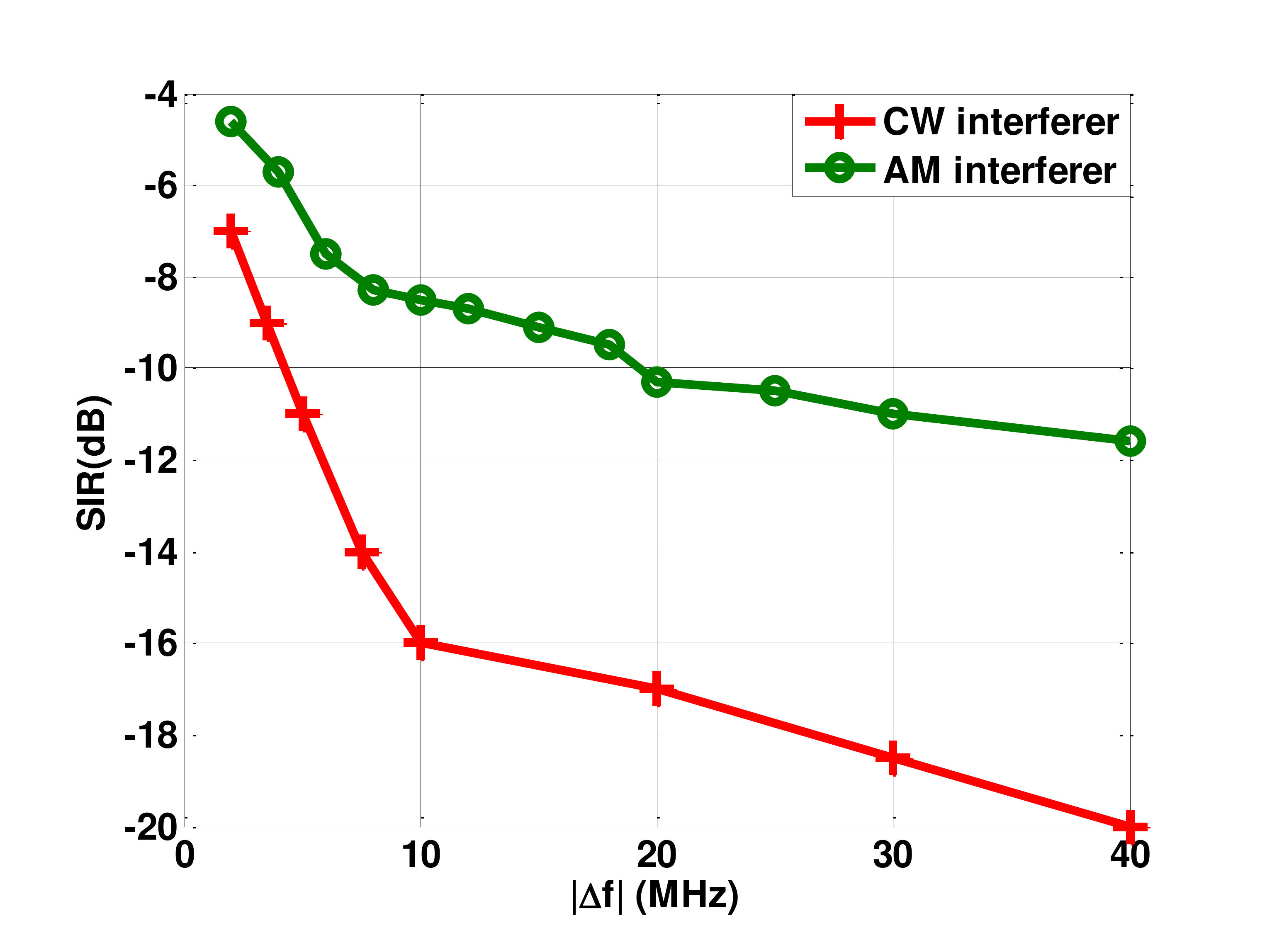}
	\caption{ Measured signal-to-interferer ratio (SIR) at different offset frequencies from carrier with CW and AM interferers at 5$\%$.}
	\label{fig:2-17}
\end{figure}
\begin{figure}[!t]
	\centering
	\includegraphics[scale=0.24]{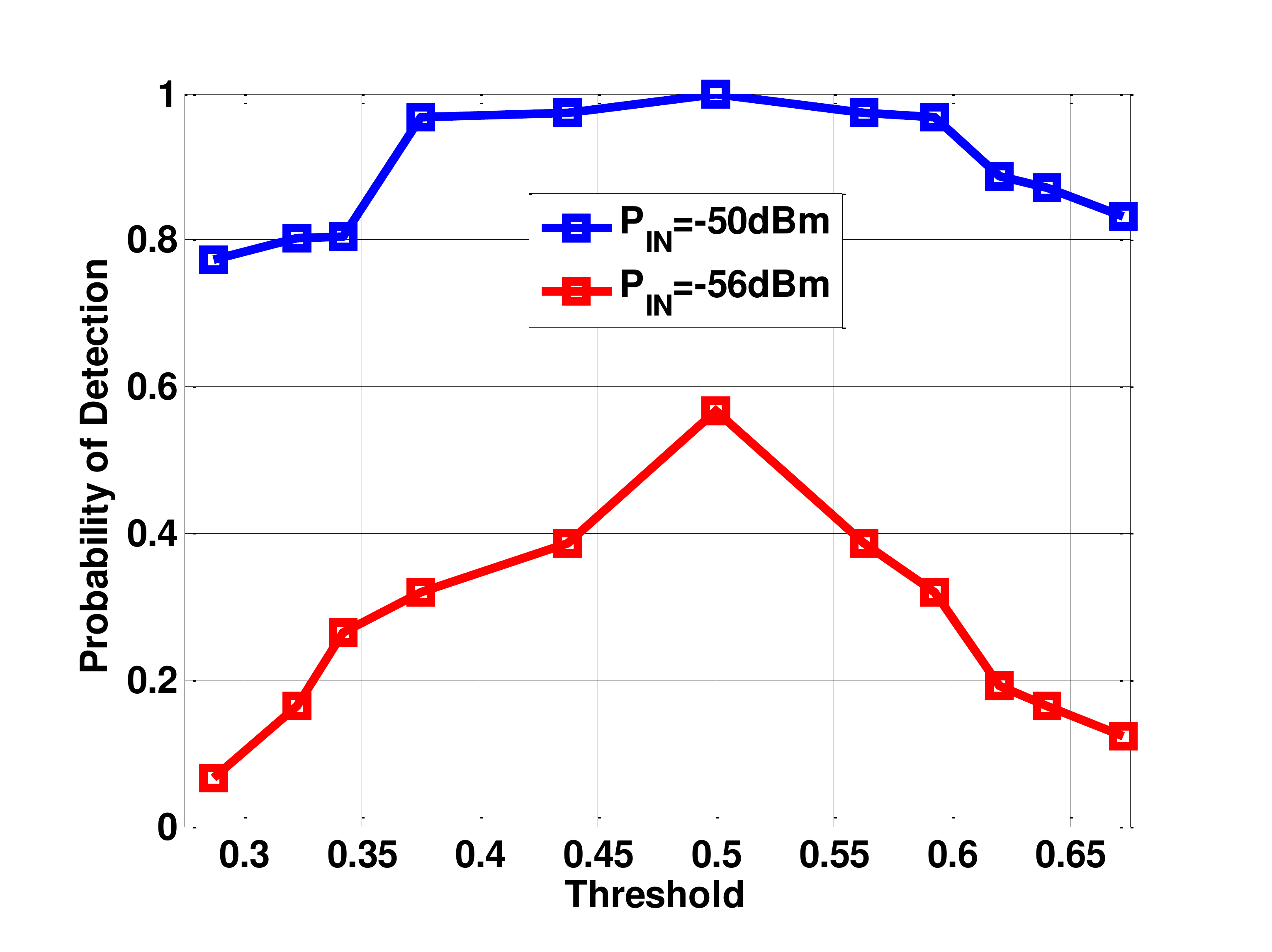}
	\caption{Measured probability of detection (PD) at $P_{IN}$=-50dBm and $P_{IN}$=-56dBm.}
	\label{fig:2-30}
\end{figure}

\begin{figure}[!t]
	\centering
	\includegraphics[scale=0.24]{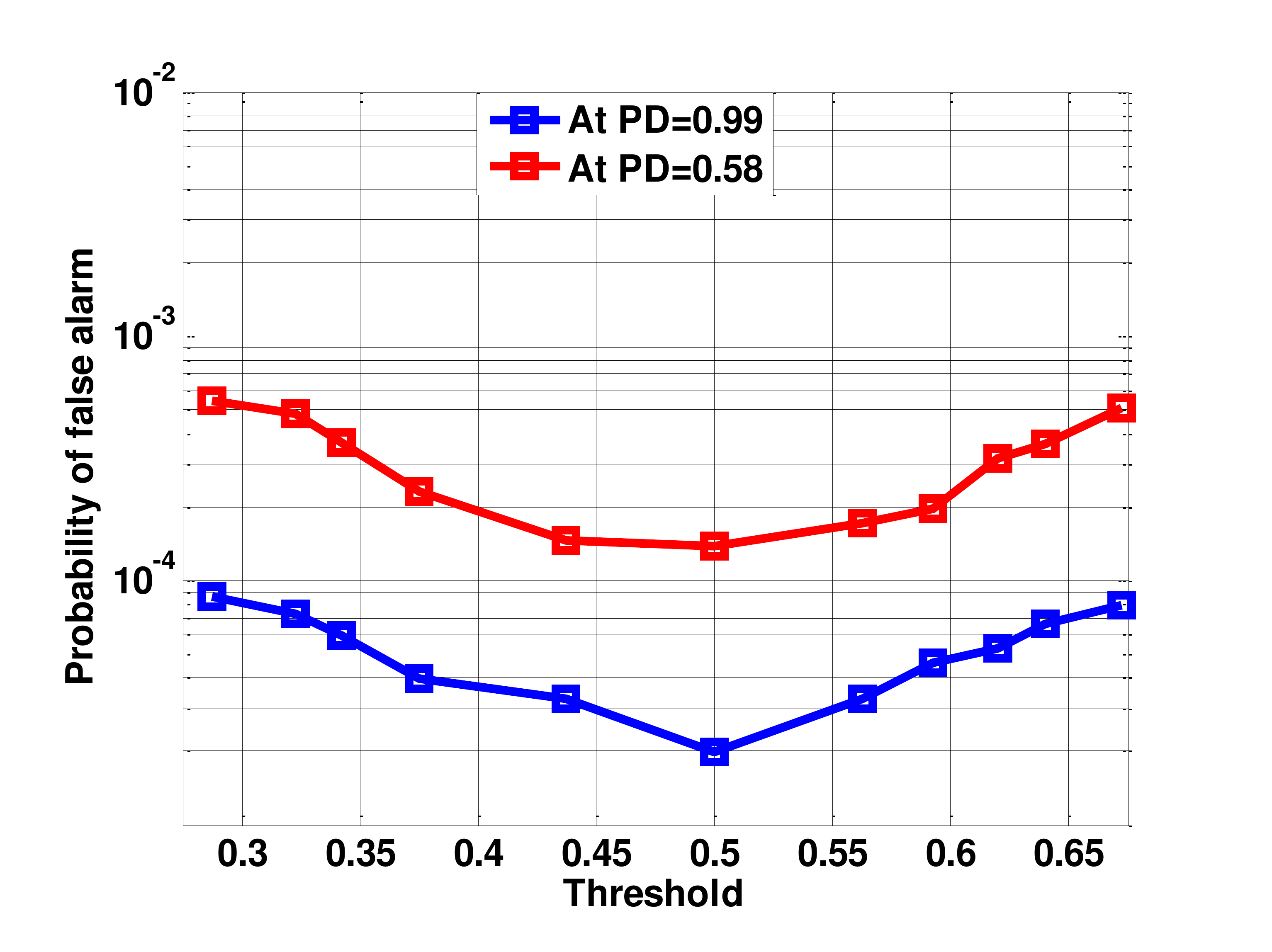}
	\caption{Measured probability of false alarm at PD=0.99 and PD=0.58.}
	\label{fig:2-20}
\end{figure}

\par
Fig. \ref{fig:2-15} shows the probability of missed detection across different input power. The signal is generated using an HP E4422B signal generator with a data pattern generated from an external PRBS-7 pulse modulator. The wake-up system achieves a sensitivity of -50dBm at a probability of missed detection of $10^{-3}$.
\begin{table*}[!t]
	\caption{Comparison of the proposed WuRx with previously published designs.}
	\label{table:T1}
	\centering
	
	\setlength{\tabcolsep}{5pt}
	\renewcommand{\arraystretch}{2}

\begin{tabular}{|l|c|c|c|c|c|c|c|}
		\hline
		& TCAS-I20 \cite{1-3-7} &   JSSC16 \cite{1-1-4} & JSSC19 \cite{1-3-6} & JSSC18 \cite{1-3-1} & CICC12 \cite{2-1-3}  & TCAS-I17 \cite{2-1-2}&  \textbf{This work} \\
		\hline
		\textbf{Technology(nm)} & 90 & 65	& 130 &	180	& 130 &	180	& \textbf{65}\\
		\hline
		\textbf{Supply voltage(V)} & 1.2	& 0.5 &	0.6/1 &	0.4	& 1.2/0.5 &	0.8	& \textbf{0.4} \\
		\hline		
		\textbf{Carrier frequency (GHz)} & 0.771 &	2.4&	0.43&	0.113&	0.403&	2.4	& \textbf{0.75}\\
		\hline		
		\textbf{Passive gain (Av) (dB)}&	12&	N/A	&27	&25	&5&	N/A&	\textbf{13}\\
			\hline
		\textbf{Data rate(kbps)}  & 2	& 10& 	0.2	& 0.3& 	12.5& 	200	& \textbf{200}\\
		\hline
		\textbf{Wake-up latency($\mu$s)}  & N/A &	N/A &	82.5 &	106	& 2.48 &	0.2	& \textbf{0.2} \\
		\hline
		\textbf{Sensitivity(dBm)}  & -46	&-97&	-76&	-69	&-45&	-50	& \textbf{-50}\\
		\hline
		\textbf{Normalized Sensitivity(dB)*}  &  -64.49&	-136&	-87.5&	-81.4&	-65.5&	-76.5&	\textbf{-76.5}\\
			\hline
		\textbf{Power ($\mu$W)}& 0.036&	99&	0.0074&	0.004&	0.116&	4.5	& \textbf{1.69}\\
		\hline
		\textbf{Energy per bit (pJ/bit)}  & 18 &	9900 &	37&	15 &	9.3	& 22.5&	\textbf{8.45} \\
		\hline
	\end{tabular}
\end{table*}
Fig. \ref{fig:2-16} shows the power breakdown of the proposed architecture when running at 200kbps at $P_{IN}$=-50dBm. The power consumption of 1.485 $\mu$A at 0.1$\%$ activity rate of the second phase translates to 7.45 pJ/bit. The primary power dissipation in the first phase is in the amplifier chain (1.088 $\mu$W). The biasing power of the Schottky diode is 288nW and the matching network consumes no power. During the operation of the second phase, an additional power of 200nW is dissipated primarily by the oscillator. The digital correlator and storage elements only consumes 13.4 nW. The two-phase approach increases operating power by 12$\%$ when compared to a single stage but dramatically reduces the probability of false alarm. 
\par
Fig. \ref{fig:2-17} shows the signal to interference ratio (SIR) of two different interferers. First, the interferer is a continuous-wave (CW) signal at an offset frequency to the wake-up signal's carrier frequency. Close-in to the carrier frequency within the bandwidth of the baseband amplifiers, the CW modulates with the carrier and interferes strongly. At an offset of 10-MHz, beyond the amplifier bandwidth, the SIR is measured at -16dB and improves with the filtering of the passive bandpass filter. The input power used in the measurement is 6-dB higher than the reported sensitivity. The second curve in the figure shows the impact of using an amplitude-modulated (AM) interferer with 5$\%$ modulation depth at bandwidth of 400kHz. Even at offset frequencies beyond the bandwidth of the amplifiers, the AM signal shows greater impact on the system compared to the CW signal leading to an SIR of only -9dB at 10-MHz offset frequency. This difference is because, after the power detection of the diodes, the CW signal appears as an offset for the correct OOK sequence. The proposed system accommodates for this offset with the proper threshold. With an AM signal, the modulation corrupts the "0" and "1" bit amplitudes. 
\par

To verify the system analysis in Section \ref{sec:model}, both the probability of detection and the probability of false alarm are measured. The probability of detection, shown in Fig. \ref{fig:2-30}, is measured at an input power of -50dBm  and -56dBm while varying the threshold voltage as a ratio of the supply voltage. The measurement results match the analysis in Section \ref{subsec:3-3-3} in which lower threshold voltages lead to higher noise of "0" to "1" bit flips and hence cause missed detection. The measurement shows the optimum threshold voltage near half $V_{DD}$. 
\begin{figure}[!t]
	\centering
	\includegraphics[scale=0.24]{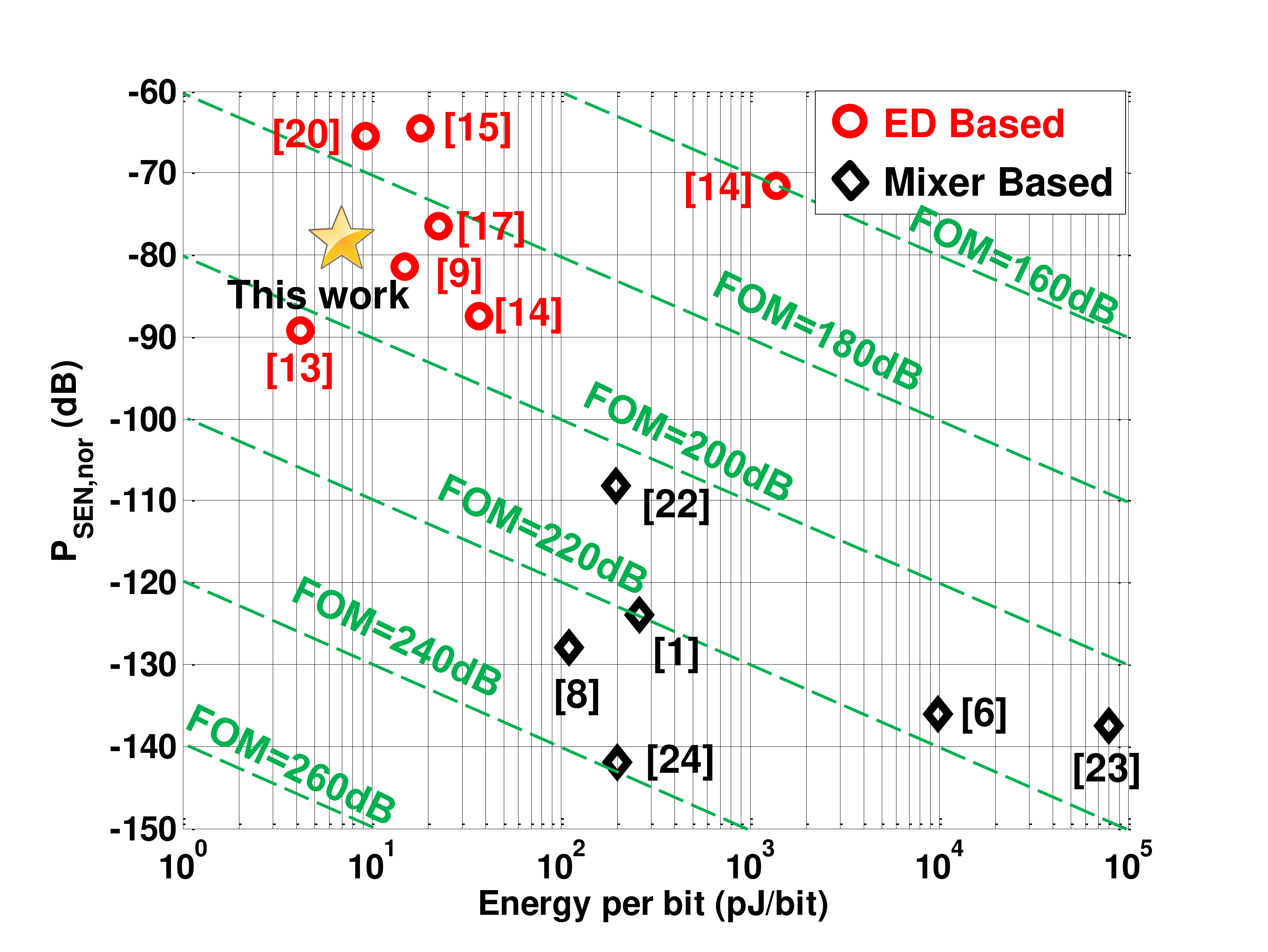}
	\caption{Scatter plot comparing the normalized sensitivity and energy per bit of previously published design and this design. Various values of the proposed FoM are shown as dashed lines.}
	\label{fig:2-18}
\end{figure}
\par
 We measured the probability of false alarm as it impacts the average power consumed in the WuRx. The probability of detection at -50dBm and -56dBm which are near the threshold of detection is used as the baseline with the probability of detection of 0.99 and 0.58 respectively. The input sequence is then replaced with a false sequence that differs from the correct target signature by 2 bits. For each threshold voltage, we repeatedly transmit the wrong sequence and counted the times a false alarm is detected.  The probability of false alarm is the count divided by $10^{6}$ which corresponds to the total number of packets sent. The results, plotted in Fig. \ref{fig:2-20}, matches the analysis results in Fig. \ref{fig:snr_thresh} showing a probability of false alarm of $<10^{-4}$. At this level, the power penalty due to false alarms is negligible.
 \par
Table \ref{table:T1} compares the performance metrics with other published designs. The energy performance of 8.45 pJ/bit compares favorably with previous publications. By using 200kbps and a short detection signature and payload (40 bits), the system wakes up with a latency of roughly only ~200$\mu$s allowing for fast response of the leaf node. 
\par
A plot comparing the proposed design with other publications is shown in Fig.\ref{fig:2-18} along with a proposed figure of merit ($FoM$). The $FoM$ is based on the following equation.
\begin{equation} \label{eq:FOM1}
FoM=-P_{SEN,nor}-10log({E/bit})
\end{equation} 
$P_{SEN,nor}(dB)$ is the normalized sensitivity in \cite{1-3-1} and is given by $P_{SEN,nor}(dB)=P_{SEN}-5log(BW_{BB})$ for a square-law detector. While this normalization properly compensates for the noise reduction from using a lower data rate, there is still a power advantage from reducing the data bandwidth. The proposed $FoM$ adjusts for the data-bandwidth dependency by using the energy per bit (E/bit) to normalize the power to different data rates. This normalization accounts for not only the digital switching ($CV_{DD}^2f_{osc}$) of the correlator but also the bias current of the Schottky diode and the amplifiers, both of which are roughly proportional to the bandwidth. The proposed $FoM$ equally weights the three main parameters of a WuRx, namely, the input sensitivity, the data rate, and the power dissipation. The proposed WuRx system achieves an $FoM$ of 187.23dB which compares favorably to prior publications as shown in the figure.

\par

\par

\section{Conclusion}\label{sec:Concsec}
In this paper, we present a two-phase WuRx that operates at a carrier frequency of 750 MHz, dissipates $<$10 pJ/bit, and achieves a low latency of 200$\mu$s. 
By using a system analysis to guide the design, we choose parameters to achieve a low probability of false alarm of the primary transceiver which in turn can lead to up to 60x energy reduction of the overall system as compared to a single-stage energy detection receiver. A Schottky-diode doubler biased at low current is used and achieves the required bandwidth while downconverting the modulated data sequence. The second phase of operation uses a data-locked startable relaxation oscillator that turns on only after the first phase detects a transition. The design consumes only 1nW/kHz, achieves $<$0.1$\%$ frequency variation across temperature, and avoids the use of an external crystal oscillator. The overall system receives up to 200kbps of data with -50dBm of sensitivity. The design is implemented in 65-nm CMOS technology and occupies an active area of 800$\mu$m$\times$350$\mu$m.

 \appendix
\renewcommand{\b}[1]{\boldsymbol{\mathrm{#1}}}
\section{Formulas}

\subsection{ED Analysis}
To model the energy detector (ED), we first make a few definitions in addition to the ones in Section II. The probability distribution of a sample $z[k]$ when $x[k]=0$ is given by
\begin{equation}
p_{0}(z)\sim N(0,\sigma^{2})
\end{equation}
and when $x[k]=1$
\begin{equation}
p_{1}(z)\sim N(1,\sigma^{2})
\end{equation}
where $N(\mu,\Lambda^{2})$
denotes a Gaussian distribution with mean $\mu$ and variance $\Lambda^{2}$.
Using a threshold $\lambda$, the probability of 1 being received when
1 is transmitted is given by \cite{detection_textbook}
\begin{equation}
p_{1C}=\int_{\lambda}^{\infty}p_{1}(z)dz=Q(\frac{\lambda-1}{\sigma})
\end{equation}
where $Q$ is the Q-function and the probability of 0 being received when 0 is transmitted is given
by
\begin{equation}
p_{0C}=\int_{-\infty}^{\lambda}p_{0}(z)dz=1-Q(\frac{\lambda}{\sigma}).
\end{equation}
Note, since we are using OOK, zero being transmitted is the same
as not transmitting any signals. The probability of noise being mistakenly
decoded as a one is given by $(1-p_{0c})$. Hence, the probability
of detection is equal to the probability of correctly decoding 1
\begin{equation}
P_{d}^{ED}=p_{1C}=Q(\frac{\lambda-1}{\sigma})
\end{equation}
However, the probability of false alarm for an ED is due to either misclassifying a 0 given $H_{0A}$ and a detection given $H_{0B}$, which can be expressed as
\begin{equation}
P_{FA}^{ED}=\frac{(1-P_{0C})P(H_{0A})+p_{1C}P(H_{0B})}{P(H_{0A})+P(H_{0B})}
\end{equation}

\subsection{Corr Analysis}
The probabilities of detection and false alarm of the correlator can be calculated from the probability of the correlator declaring a match under the $H_{0A}$, $H_{0B}$, and $H_{1}$ using the following relations
\begin{equation}
P_{D}^{Corr}=P(\text{match}|H_{1})
\end{equation}
\begin{equation}
\begin{aligned}
P_{FA}^{Corr}  =&p(\text{match}|H_{1}\cup H_{0A})\\
 =&p(\text{match}|H_{0A})\frac{P(H_{0A})}{P(H_{0A})+P(H_{0B})}\\ &+p(\text{match}|H_{0B})\frac{P(H_{0B})}{P(H_{0A})+P(H_{0B})}
 \end{aligned}
 \label{eq:pfa_comb}
\end{equation}
The derivation of $p(\text{match}|H_{0A})$, $p(\text{match}|H_{0B})$, and $p(\text{match}|H_{1})$ requires the consideration of different combinations of the signature $\b{s}$ and the received bits, and hence, is more involved than the energy detector.  We start by discussing the expression for the simple case where all bits need to be decoded correctly ($l=0$) and then discussing the general case where a match is declared despite having a mistake in $l\geq0$ bits.
\subsubsection{l=0}
Assuming a threshold $\lambda$ yielding probability $P_{1C}$ and
$P_{0C}$ same as previously defined, the probability of a sequence
of length $L$ and $d$ ones to be received correctly, i.e. all bits
match, is given by
$
P(\text{match}|H_{1})=(P_{1C})^{d}(P_{0C})^{L-d}
$.
The probability of an error given $H_{0A}$ is similar. Since nothing is being transmitted, $d$ erroneous ones need to be decoded
with nothing transmitted. Hence, we replace $P_{1C}$ with $(1-P_{0C})$
to obtain
$
P(\text{match}|H_{0A})=(1-P_{0c})^{d}(P_{0c})^{L-d}
$.

The probability $P(\text{match}|H_{0B})$ is the probability of a sequence $\b{s}'$ to have errors in detection such that it ends up decoded as $\b{s}$. For any sequence $\b{s}'\neq \b{s}$, we define
$j=\sum_{k=1}^{L}s[k]\&(\sim s'[k]$) as the number of ones that differ
from $\b{s}$ and $i=\sum_{k=1}^{L}(\sim s[k])\&s'[k]$ as the number of zeros that differ
from $\b{s}$,
where $\sim$ denotes logical NOT operator and $\&$ logical AND operator.
The number of sequences that differ in $j$ ones and $i$ zeros from
$\b{s}$ is given by $C(i,j)=\left(\begin{array}{c}
L-d\\
i
\end{array}\right)\left(\begin{array}{c}
d\\
j
\end{array}\right)$. The probability of a match to occur is given by 
\begin{equation}
P(\text{match}|i,j)=(1-P_{1C})^{i}P_{0C}^{L-d-i}(1-P_{1C})^{i}P_{0C}^{L-d-i}
\end{equation}
which is the probability of having $i$ zeros and $j$ ones to flip
while the remaining bits to be decoded correctly. When assuming that all
$2^{L}-1$ sequences in $H_{0B}$ (excluding $\b{s}$) are equally probable, the probability
of a match  given $H_{0B}$ is  
\begin{align*}
p(\text{match}|H_{0B}) & =\frac{1}{2^{L}-1}\sum_{j=0}^{d}\sum_{i=0,i+j>0}^{L-d}C(i,j)P(\text{match}|i,j)\\
\end{align*}

\subsubsection{ l $\geq$ 0}

We derive the probability of a match under the three hypotheses when we consider
sequences that differ from the target, $\b{s}$, by at most $l$ bits 
as a match. Let $g$ be defined as the number of differences between
a received sequence $\hat{\b{s}}$ and the target such that $g=\sum_{i=0}^{L}\hat{s}[k]\neq s[k]$.

The probability of receiving a sequence $\b{s}'$ which differs in exactly
$k$ bits from $\b{s}$ given that $\b{s}$ was transmitted is given by 
\begin{align*}
P& (g=k\text{ }|H_{1})= & \\ &\sum_{\begin{array}{c}
	i+j=k,\\0\leq i,j\leq L
	\end{array}}\left(\begin{array}{c}
d\\
i
\end{array}\right)(P_{1C})^{d-i}\left(\begin{array}{c}
L-d\\
j
\end{array}\right)(P_{0C})^{L-d-j}
\end{align*}
which is the probability of decoding $d-i$ ones correctly and $L-d-j$
zeros correctly for all combinations satisfying $i+j=k$. The probability
of a match given $H_1$, hence, is the summation over all possibilities of having
$k\leq l$  mistakes given by $P(\text{match}|H_{1})  =\sum_{k=0}^{l}P(g =k|H_{1})$

The probability of a match with $l$ bits different under $H_{0A}$ is
similar to $P(\text{match}|H_{1})$, except that $P_{1C}$ is replaced with
$(1-P_{0c})$ similar to what we did for $l=0$.

A match  occurs under $H_{0B}$  when a sequence
$\b{s}'\neq \b{s}$ is transmitted and a sequence $\hat{\b{s}}$ is received which
differs by at most $l$ bits from $s$. Let $\hat{\b{s}}(n)$ denote a received sequence having $n$ ones,
the probability of $\b{s}'$ being received as $\hat{\b{s}}$ is given by
\begin{equation*}
P(\b{s}'\text{ rec }\hat{\b{s}}(n))=\frac{1}{2^{L}-1}\sum_{j=0}^{n}\sum_{i=0}^{L-n}(C(i,j)-r)P(\text{match}|i,j)
\end{equation*}
where the value of $r$ accounts for the fact that the sequence $\b{s}$ is
never transmitted under $H_{0B}$ and is given by $
r=\begin{cases}
1 & i+j=k\\
0 & OW
\end{cases}
$ . When
the sequence $\hat{\b{s}}$ differs from $\b{s}$ by exactly $k$ bits, the
number of ones in $\hat{\b{s}}$ can vary from $n=d-k$ to $n=d+k$. The
number of sequences that differ from $\b{s}$ (which has $d$ ones and
$L-d$ zeros) in $k$ bits and have $n$ zeros is given by $C'(k,n).$
The probability of a sequence having exactly $k$ mistakes being decoded
as $\b{s}$ is given by
\begin{equation}
P(\text{match }g=k)=\sum_{n=d-k}^{d+k}C'(k,n)P(\b{s}'\text{ rec }\hat{\b{s}}(n))
\end{equation}
where $C'(k,n)=$
\begin{multline*}
\begin{cases}
\left(\begin{array}{c}
L-d\\
h+f
\end{array}\right)\left(\begin{array}{c}
d\\
h
\end{array}\right), &
\begin{array}{c} f\geq0,L-d\geq h+f, \\ d\geq h,h\geq0,\\ 2h+|f|=k\end{array}\\
\left(\begin{array}{c}
L-d\\
h
\end{array}\right)\left(\begin{array}{c}
d\\
h-f
\end{array}\right) & \begin{array}{c}
f<0,L-d\geq h,\\ d\geq h-f, \\h\geq0,2h+|f|=k
\end{array}\\
0 & OW
\end{cases}
\end{multline*}
and  $h=\left\lfloor \frac{k-|f|}{2}\right\rfloor $ and $f=n-d$.
The probability of match given $H_{0B}$ is given by
\begin{equation*}
P(\text{match}|H_{1})=P(\text{match }g\leq l)=\sum_{k=0}^{L}P(\text{match }g=k)
\end{equation*}
The derived expressions were verified against simulations in the paper.%

\subsection{OOK MF Analysis}

\begin{figure}[t!]
	\centering
	\includegraphics{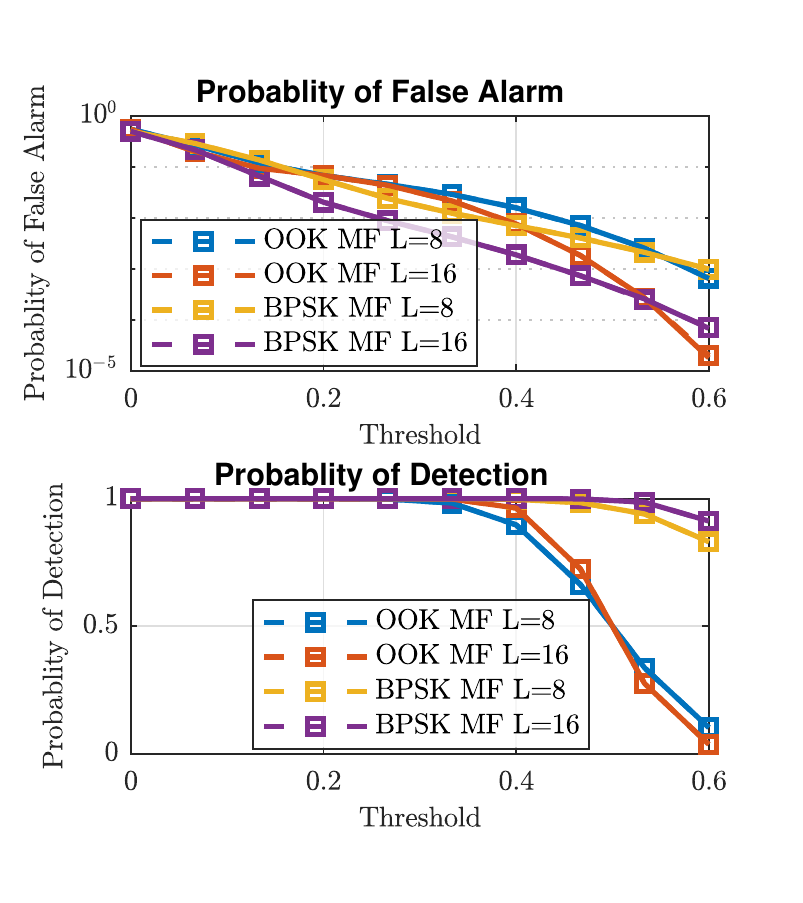}
	\caption{Probability of false alarm and detection as a function of the threshold for a matched filter detector. Both simulation results (solid lines) and analytical results (dashed lines with markers) overlap.}
	\label{fig:mf_L_thresh}
\end{figure}

Since $\b{s}$ has $L-d$ zeros and we multiply the input by $\b{s}$, the zeros
do not affect our decision. So, $\eta$ is the sum of $d$ random
Gaussian random variables. Under $H_{1}$, these variable have a mean
equal to $d$ making
$
\eta\ |H_{1}\sim N(d,d\sigma^{2})
$
leading to a probability of detection equal to
\begin{equation}
P_{d}^{MF}=p(\eta>\lambda^{MF}|H_{1})=Q(\frac{\lambda^{MF}-d}{\sqrt{d}\sigma})
\end{equation}
Under $H_{0A}$, no signal was transmitted and the the mean is equal to zero making
$
\eta\ |H_{0A}\sim N(0,d\sigma^{2})
$
leading to 
\begin{equation}
P(\text{pos}|H_{0A})=p(\eta>\lambda^{MF}|H_{0A})=Q(\frac{\lambda^{MF}}{\sqrt{d}\sigma})
\end{equation}
Under $H_{0B}$, the mean for a sequence having $i$ ones, is $i$.
 This is given by
\begin{align}
P(\text{pos}|H_{0B}) &= p(\eta>\lambda^{MF}|H_{0B}) \\&=\frac{1}{2^{L}-1}\sum_{i=0}^{d}\left(C''(i)Q\left(\frac{\lambda^{MF}-i}{\sqrt{d}\sigma}\right)\right)
\end{align}
where $C''(i)$ is the number of sequences having at least $i$  ones matching the ones of $s$ and it is given by 
$
C''(i)=\left(\begin{array}{c}
d\\
i
\end{array}\right)2^{L-d}-r 
$
where $r$ equals 1, when $i=d$ and zero otherwise to account for the fact that $\b{s}$ is never sent under $H_{0B}$. The derived expressions are verified by simulations in Fig.~\ref{fig:mf_L_thresh}. The probability of false alarm can be calculated from $P(\text{pos}|H_{0A})$ and $P(\text{pos}|H_{0B})$ similar to (\ref{eq:pfa_comb}).

\subsection{BPSK MF Analysis}
In the case of the BPSK matched filter, $\eta$ is the sum of $L$ random
Gaussian random variables. Under $H_{1}$, these variable have a mean
equal to $L$ and variance equal to $L\sigma^{2}$ making
$
\eta\ |H_{1}\sim N(L,L\sigma^{2})
$
leading to a probability of detection equal to
\begin{equation}
P_{d}^{MF}=p(\eta>\lambda^{MF}|H_{1})=Q(\frac{\lambda^{MF}-L}{\sqrt{L}\sigma})
\end{equation}
Under $H_{0A}$, the mean is equal to zero, and hence the  distribution is
given by
$
\eta\ |H_{0A}\sim N(0,L\sigma^{2})
$
making the probability of declaring a positive
\begin{equation}
P(\text{pos}|H_{0})=P(\eta>\lambda^{MF}|H_{0A})=Q(\frac{\lambda^{MF}}{\sqrt{L}\sigma})
\end{equation}
Under $H_{0B}$, the mean for a sequence having $i$ mistakes, is $L-2i$
where each wrong bit will decrease the expected value of the statistic
by two. Since we have $\left(\begin{array}{c}
L\\
i
\end{array}\right)$ of each of these sequences, the probability of exceeding the threshold
is given by
\begin{align}
P(\text{pos}|H_{0B}) &= p(\eta>\lambda^{MF}|H_{0B}) \\
&=\frac{1}{2^{L}-1}\sum_{i=1}^{L}(\left(\begin{array}{c}
L\\
i
\end{array}\right))Q(\frac{\lambda^{MF}-L-2i}{\sqrt{L}\sigma})
\end{align} 
The probability of false alarm can be calculated from  $P(\text{pos}|H_{0A})$ and $P(\text{pos}|H_{0B})$  similar to (\ref{eq:pfa_comb}). The derived expressions are verified by simulations in Fig.~\ref{fig:mf_L_thresh}.

\end{document}